\documentclass[a4paper,fleqn,usenatbib]{mnras}
\usepackage{newtxtext,newtxmath}
\usepackage[T1]{fontenc}
\usepackage{ae,aecompl}

\usepackage{graphicx}	
\usepackage{amsmath}	
\usepackage{amssymb}	

\usepackage{color,soul}

\definecolor{lightblue}{rgb}{.70,.95,1}
\sethlcolor{lightblue} 

\usepackage{xspace} 
\renewcommand{\AA}{\normalfont\r{A}\xspace} 

\newcommand{\teff}{\ensuremath{T_{\mathrm{eff}}}\xspace}
\newcommand{\kms}{\ensuremath{\rm{km}\,s^{-1}}\xspace}
\newcommand{\logg}{\ensuremath{\log g}\xspace}
\newcommand{\feh}{\rm{[Fe/H]}\xspace}
\newcommand{\cfe}{\rm{[C/Fe]}\xspace}

\newcommand{\alphafe}{\rm{[\ensuremath{\alpha}/Fe]}\xspace}
\newcommand{\Gaia}{\textit{Gaia}\xspace}

\title[PIGS II: survey description]{The Pristine Inner Galaxy Survey (PIGS) II: Uncovering the most metal-poor populations in the inner Milky Way\thanks{based on observations made with the Canada-France-Hawaii Telescope (CFHT) and the Anglo-Australian Telescope (AAT)}}
 
\author[A. Arentsen et al.]{
Anke Arentsen,$^{1}$\thanks{E-mail: aarentsen@aip.de}
Else Starkenburg,$^{1}$
Nicolas F. Martin,$^{2,3}$
David S. Aguado,$^{4}$
\newauthor
Daniel B. Zucker,$^{5}$ 
Carlos Allende Prieto,$^{6,7}$
Vanessa Hill,$^{8}$
Kim. A. Venn,$^{9}$
\newauthor
Raymond G. Carlberg,$^{10}$
Jonay I. Gonz\'alez Hern\'andez,$^{6,7}$
Lyudmila I. Mashonkina,$^{11}$
\newauthor
Julio F. Navarro,$^{5}$
Rub\'en S\'anchez-Janssen,$^{12}$
Mathias Schultheis,$^{10}$
\newauthor
Guillaume F. Thomas,$^{13}$
Kris Youakim,$^{1}$
Geraint F. Lewis,$^{14}$
Jeffrey D. Simpson,$^{15}$
\newauthor
Zhen Wan,$^{14}$
Roger E. Cohen,$^{16}$
Doug Geisler,$^{17,18,19}$
Julia E. O'Connell$^{17}$ \\
\\
Affiliations are listed after the references
}

\date{Accepted XXX. Received YYY, in original form ZZZ}

\pubyear{2020}

\begin{document}
\label{firstpage}
\pagerange{\pageref{firstpage}--\pageref{lastpage}}
\maketitle

\begin{abstract}
Metal-poor stars are important tools for tracing the early history of the Milky Way, and for learning about the first generations of stars. Simulations suggest that the oldest metal-poor stars are to be found in the inner Galaxy. Typical bulge surveys, however, lack low metallicity ($\feh < -1.0$) stars because the inner Galaxy is predominantly metal-rich. The aim of the Pristine Inner Galaxy Survey (PIGS) is to study the metal-poor and very metal-poor (VMP, $\feh < -2.0$) stars in this region. In PIGS, metal-poor targets for spectroscopic follow-up are selected from metallicity-sensitive $CaHK$ photometry from the CFHT. This work presents the $\sim 250$ deg$^2$ photometric survey as well as intermediate-resolution spectroscopic follow-up observations for $\sim 8000$ stars using AAOmega on the AAT. The spectra are analysed using two independent tools: ULySS with an empirical spectral library, and FERRE with a library of synthetic spectra. The comparison between the two methods enables a robust determination of the stellar parameters and their uncertainties. We present a sample of 1300 VMP stars -- the largest sample of VMP stars in the inner Galaxy to date. Additionally, our spectroscopic dataset includes $\sim1700$ horizontal branch stars, which are useful metal-poor standard candles. We furthermore show that PIGS photometry selects VMP stars with unprecedented efficiency: 86\%/80\% (lower/higher extinction) of the best candidates satisfy $\feh < -2.0$, as do 80\%/63\% of a larger, less strictly selected sample.
We discuss future applications of this unique dataset that will further our understanding of the chemical and dynamical evolution of the innermost regions of our Galaxy.

\end{abstract}



\section{Introduction}

The oldest, metal-poor stars that still exist today formed in relatively pristine environments and are unique probes of the conditions in the early universe. These stars carry the imprint of the first supernovae and are local windows into the first episodes of star formation. Studying their chemistry in detail advances our understanding of the first generations of stars and their supernovae, and, combined with kinematic information, metal-poor stars provide clues to the early formation history of the Milky Way and its different components \citep[see reviews by][]{freeman02, tolstoy09, frebelnorris15}.
 
The $\Lambda$ cold dark matter simulations of \citet{diemand05} predict that, if any of the very first stars have survived to the present day, $30-60\%$ of them would be within 3 kpc of the Galactic centre. Similarly, the simulations of \citet{tumlinson10} suggest that metal-poor stars below \feh\footnote{[X/Y] $ = \log(N_\mathrm{X}/N_\mathrm{Y})_* - \log(N_\mathrm{X}/N_\mathrm{Y})_{\odot}$, where the asterisk subscript refers to the considered star, and N is the number density.} $=-3.5$ (all formed at redshifts $> 15$) would be most frequently found within the bulge, because dark matter haloes grow inside-out. These stars are not necessarily born in the bulge, but they have ended up in the central regions because of their dynamical history. Using hydrodynamical simulations, \citet{starkenburg17a} also find that the fraction of stars that are both metal-poor (\feh $< -2.5$) and old (formed within the first 0.8 Gyr) is highest towards the centres of galaxies, even though the relative fractions of just metal-poor stars or just old stars increase with galactic radius. Hence, a clear picture emerges that the central part of our Galaxy is a privileged location to look for the oldest and most metal-poor stars. 

A complication in studying old, metal-poor stars in the Galactic bulge is that
the overwhelming majority of stars are metal-rich ($\feh > -0.5$). This metal-rich component has been studied in detail in previous surveys such as the GIRAFFE Inner Bulge Survey \citep[GIBS,][]{zoccali14}, the Bulge Radial Velocity Assay \citep[BRAVA, ][]{kunder12}, the Abundances and Radial velocity Galactic Origins Survey \citep[ARGOS,][]{ness12,freeman13} and the Apache Point Observatory Galactic Evolution Experiment \citep[APOGEE, e.g.][]{schultheis17,garciaperez18, zasowski19}. The general conclusion from these previous studies is that the Galactic bulge is a rotating, boxy/peanut-shaped pseudo-bulge, with little room for a pressure-supported, classical bulge \citep[e.g.][]{howard09, hill11, shen10, ness13b}. If there is a classical component, it can only contain a fraction of the mass of the entire bulge and it is likely more dominant at lower metallicities. 

The kinematics of the metal-poor tail of the bulge are still largely unexplored, with the exception of some RR Lyrae studies \citep[e.g.][]{dekany13,kunder16}. The ARGOS survey of bulge red clump stars, with a survey strategy aimed to be metallicity-unbiased, found only 16 out of $14\,150$ stars with $-2.6 < \feh < -2.0$, and only 84 stars with $\feh < -1.5$ \citep{ness13a}. This emphasises that targeted surveys are necessary for the discovery of large samples of metal-poor stars in the inner Galaxy. 

There have been several efforts to find very metal-poor (\feh $ < -2.0$) stars in the inner Galaxy using photometry. \citet{caseyschlaufman15} employed a combination of infrared 2MASS and WISE photometry. These authors found and studied three \feh~$< -2.7$ stars with high-resolution spectroscopy, but it was limited to bright stars only (V~$<14$). Another study by \citet{koch16}, used a filter focused on the Ca~II~K line to select candidate metal-poor stars. They found four stars with $-2.7 < \feh < -2.3$, among them the first carbon-enhanced metal-poor star in the inner Galaxy. 

So far, the largest effort to search for metal-poor stars in the inner Galaxy has been the Extremely Metal-poor BuLge stars with AAOmega (EMBLA) survey \citep{howes14,howes15,howes16}, which used SkyMapper photometry. The SkyMapper filter set includes a narrow-band $v$ filter containing the Ca H\&K lines. These lines are good metallicity indicators, and in combination with broad-band photometry, metal-poor stars can be selected efficiently using this $v$ filter \citep{wolf18}. The EMBLA team has performed low-resolution follow-up spectroscopy of photometrically selected candidates and identified $\sim 150$ stars with \feh $< -2.5$ (L. Howes, priv. comm.). From this sample, 37 stars were selected for high-resolution spectroscopic follow-up, and of those, 30 have \feh $\leq -2.5$, and 9 have \feh $\leq -3.0$. These stars are the first known extremely metal-poor stars in the inner Galaxy. The recent Chemical Origins of Metal-poor Bulge Stars (COMBS) survey also analysed stars selected using SkyMapper photometry. Their high-resolution spectroscopic sample contains five stars with $\feh < -2.0$ \citep{lucey19}. 

A striking difference between the metal-poor EMBLA stars in the inner Galaxy and typical Galactic halo stars is that only a small fraction of EMBLA stars appear to be enhanced in carbon \citep{howes15, howes16}. In the Galactic halo, the number of stars with enhanced carbon increases with decreasing metallicity. These carbon-enhanced metal-poor (CEMP) stars comprise 15--20\% of stars with $\feh < -2$ and up to 70\% of stars with $\feh < -4$ \citep{yong13a, placco14}. One of the reasons for the lack of CEMP stars in EMBLA could be that the survey is biased against carbon-rich stars. The SkyMapper $v$ filter includes a region of the spectrum where a CN-band can be present in carbon-enhanced stars, which makes such stars appear more metal-rich and fail the selection for follow-up \citep{dacosta19}. 

The low number of known metal-poor stars in the bulge, the apparent lack of CEMP stars in the inner Galaxy, the possibility to uncover the oldest, metal-poor stars in the Milky Way, and the opportunity to study the metal-poor tail of the Galactic bulge are all good reasons to expand the search for metal-poor stars in the inner Galaxy. The Pristine Inner Galaxy Survey (PIGS) aims to build an unprecedentedly large sample of old, metal-poor stars for detailed chemo-dynamical studies. The Pristine survey \citep[][hereafter S17b]{starkenburg17b} is very efficient at uncovering metal-poor stars photometrically. It has reported success rates of 56\% and 23\% from spectroscopic follow-up studies for finding stars in the Galactic halo with $\feh < -2.5$ and $\feh < -3.0$, respectively \citep{youakim17, aguado19, venn20}. In the Pristine survey, a narrow-band Ca H\&K filter on MegaCam at the Canada-France-Hawaii Telescope (CFHT) is used to provide photometric metallicities down into the extremely metal-poor regime. 
The $CaHK$ photometry is combined with available broadband photometry for colour/temperature information to compute these photometric metallicities. The Pristine filter is narrower than the SkyMapper filter, and should be less biased against CEMP stars (see Figure~2 in S17b). 

\begin{figure*}
\centering
\includegraphics[width=0.45\hsize,trim={0.0cm 0.0cm 0.0cm 0.0cm}]{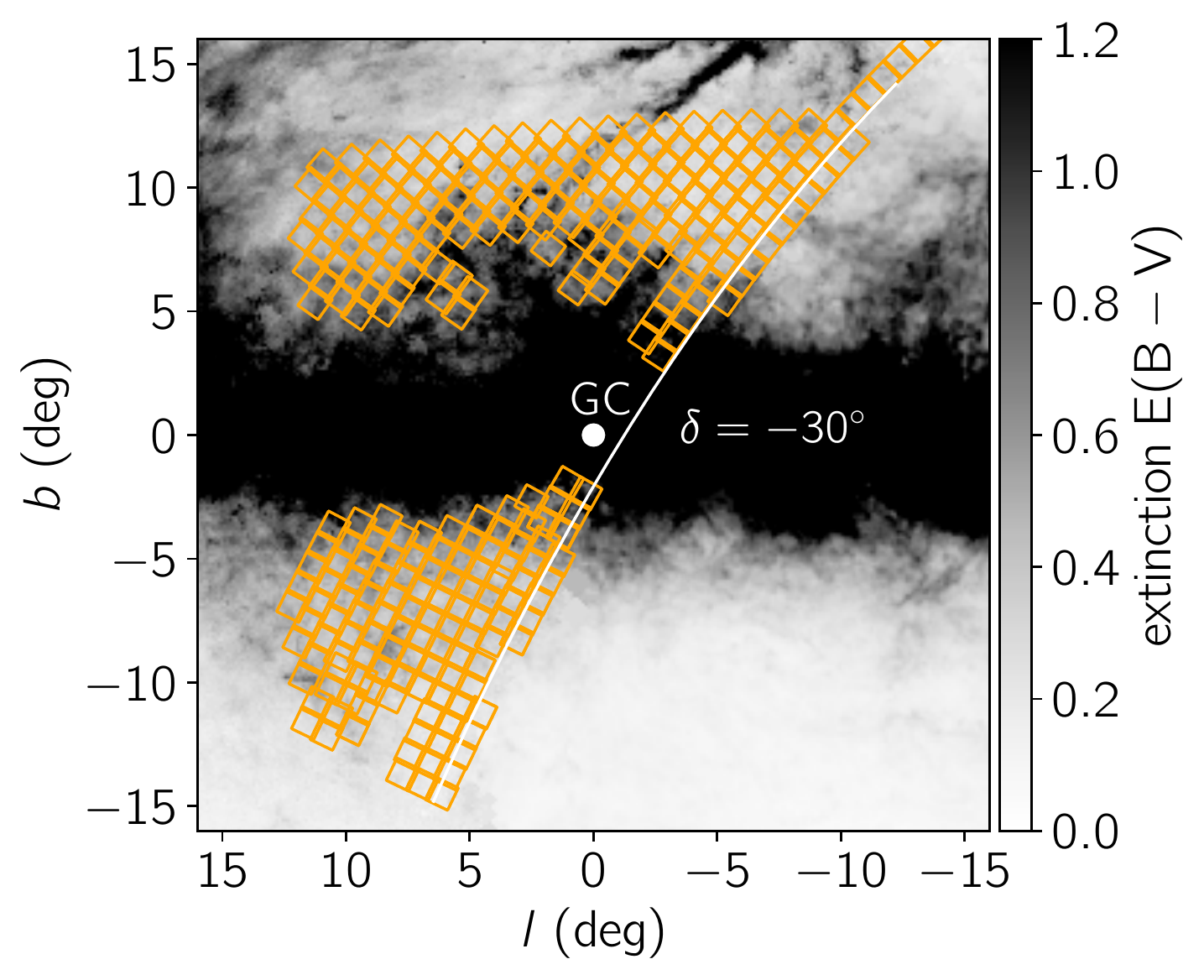}
\includegraphics[width=0.45\hsize,trim={0.0cm 0.0cm 0.0cm 0.0cm}]{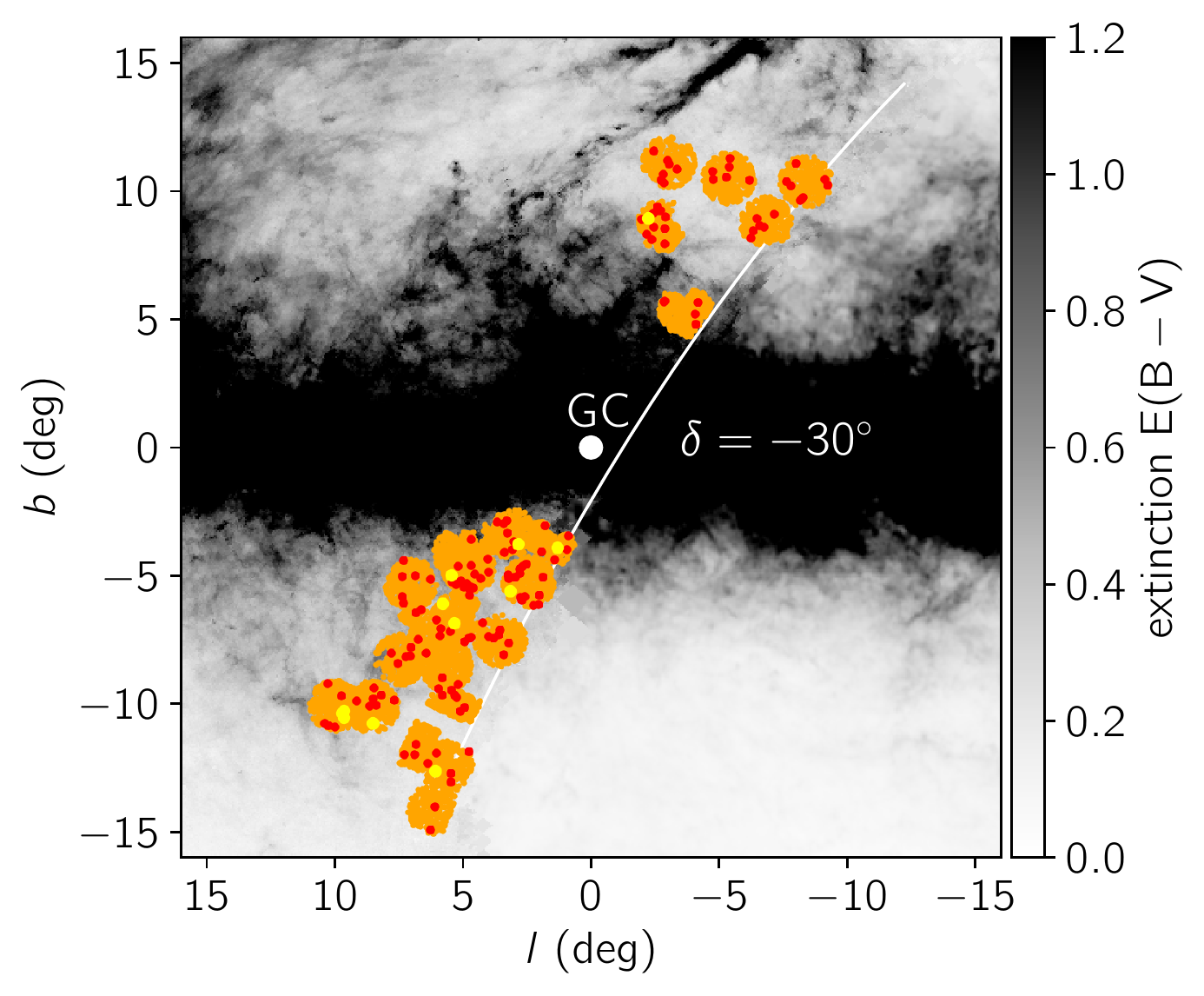}
\caption{PIGS footprints, on top of the extinction map from \citet{green18} (where available and reliable, which is above a declination of $-30^{\circ}$ and at $|b| > 2^{\circ}$) and \citet{schlegel98} (elsewhere). A truncation of the colour bar at E(B$-$V) = 1.2 is set for clarity. Left: photometric footprint, where each square is one MegaCam field. Right: spectroscopic footprint, located within the photometric footprint from 2017 and 2018. Stars with $\feh < -2.5$ are marked in red, and those with $\feh < -3.0$ in yellow.}
    \label{fig:footprint}
\end{figure*}

Compared to the halo, the inner Galaxy has a much higher density of (mainly metal-rich) stars. It is therefore significantly more crowded, there is a larger foreground population, and there is much more (inhomogeneous) reddening due to the presence of dust. These challenges, together with the unique science case, have led to a change in strategy between the main Pristine survey and PIGS, which we will describe in this paper. The first scientific results from PIGS were published in a Letter in this journal \citep{arentsen20}, and it focused solely on the kinematics of the metal-poor ($\feh < -1.0$) tail of the inner Galaxy using a sample of several thousand low- and intermediate resolution spectroscopic observations. 

In this work, we present the PIGS data collection and analysis in detail and focus on our results and success rates to find and analyse the most metal-poor stars in the inner Galaxy. The photometric $CaHK$ observations and selection of metal-poor candidates are described in Section~\ref{sec:phot}.  
The analysis of low-/medium resolution spectroscopic follow-up observations are described in Section~\ref{sec:spec}, the results for the stellar parameters are discussed in Section~\ref{sec:speccomp}, and the survey performance is determined in Section~\ref{sec:performance}. A summary of our PIGS survey and future plans are provided in Section~\ref{sec:conclusion}. 

\section{Photometric observations}\label{sec:phot}

In this section, the photometric PIGS data is presented. We discuss the data reduction and calibration, and the target selections for the spectroscopic follow-up.

With MegaCam on CFHT, we have obtained $\sim$250 square degrees of metallicity-sensitive $CaHK$ photometry in the Galactic bulge region. The footprint covers a region of $|l|$ and $|b| \lesssim 12^\circ$ above a declination of $-30^\circ$ (because we observe from the Northern hemisphere), and we limited it to areas with E(B$-$V) $\lesssim 0.7$ to avoid regions with too much extinction. A pilot program was initiated in 2017A, and extended to a larger program in 2018A and 2019A. The final photometric PIGS footprint is shown in the left panel of Figure~\ref{fig:footprint}. A small overlap between neighbouring MegaCam fields is included for calibration purposes.  

\subsection{Data reduction and calibration}\label{sec:photreduction}
\subsubsection{Reduction}

The PIGS MegaCam images are processed similarly to the Pristine images, as presented in S17b. All downloaded images are pre-processed by CFHT with the Elixir pipeline, whose role is to remove instrumental signatures from the data \citep{elixir}. 
The images were processed with a version of the Cambridge Astronomical Survey Unit pipeline \citep[CASU,][]{casu} tailored to MegaCam images. The astrometry is performed on a CCD to CCD basis and using 2MASS \citep{2mass} for reference. While we could also use the \Gaia DR2 \citep{gaia16, gaia18} astrometry, the shallower 2MASS data works well for the densely populated regions of the survey and has the benefit of converging more quickly on a good astrometric solution, accurate at the $0.1''$ level. This accuracy is more than enough to cross-match the sources to \Gaia DR2 positions, and we use the \Gaia positions for the stars throughout this paper. With the CASU pipeline, aperture photometry is carried out on the individual CCDs to build a catalogue of sources that are at least 5$\sigma$ above the background threshold. 

The following quality cuts are made on the \Gaia cross-matched catalogue to create our final non-calibrated catalogues: 
\begin{itemize}
	\item casu\_flag = $-1$ or casu\_flag = $-2$ 
	\item \Gaia phot\_variable\_flag $\neq$ VARIABLE 
	\item \Gaia phot\_bp\_mean\_mag and phot\_rp\_mean\_mag $> 0$
	\item \Gaia duplicated\_source = false 
\end{itemize}

The first cut only keeps point-sources from the CASU $CaHK$ photometry data reduction. The other three are quality cuts on the \Gaia data, removing variable stars, stars that do not have $BP$ and/or $RP$ photometry, and stars that are flagged to be duplicated sources. The last flag proved to be unnecessary in hindsight. The full catalogue consists of more than 34 million objects with a \Gaia DR2 counterpart (there are fewer unique stars; some stars have multiple entries due to the overlap between fields).

\begin{figure*}
\centering
\includegraphics[width=0.8\hsize,trim={0.0cm 0.0cm 0.0cm 0.5cm}]{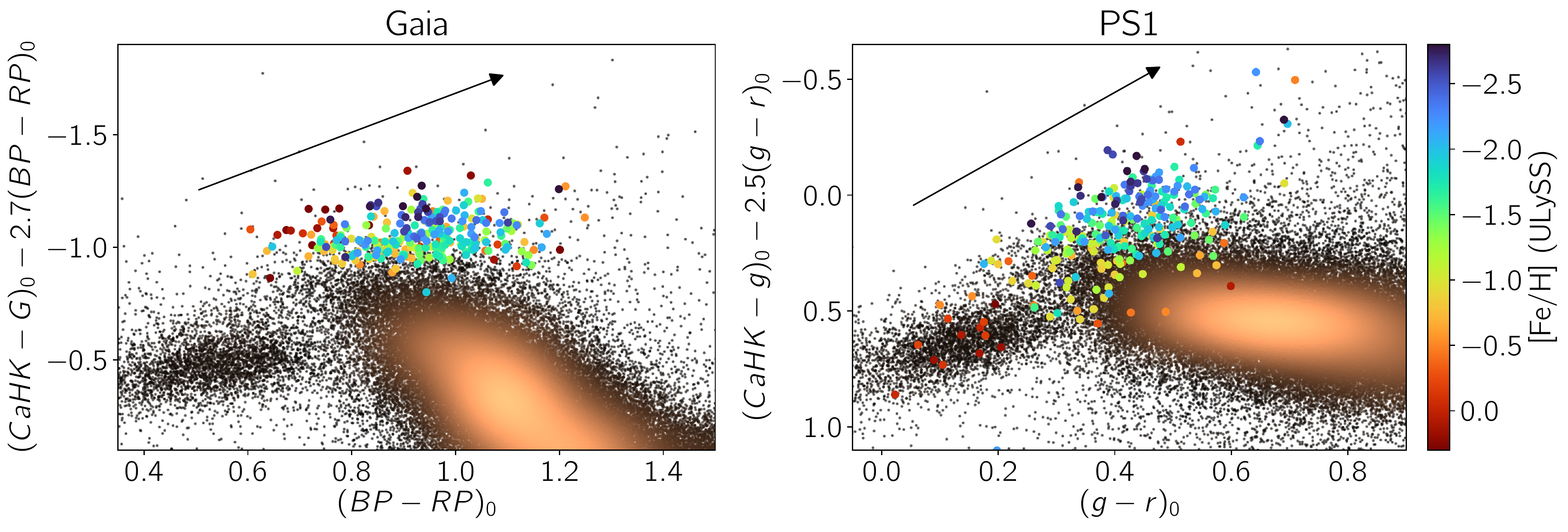}
\includegraphics[width=0.8\hsize,trim={0.0cm 0.0cm 0.0cm 0.0cm}]{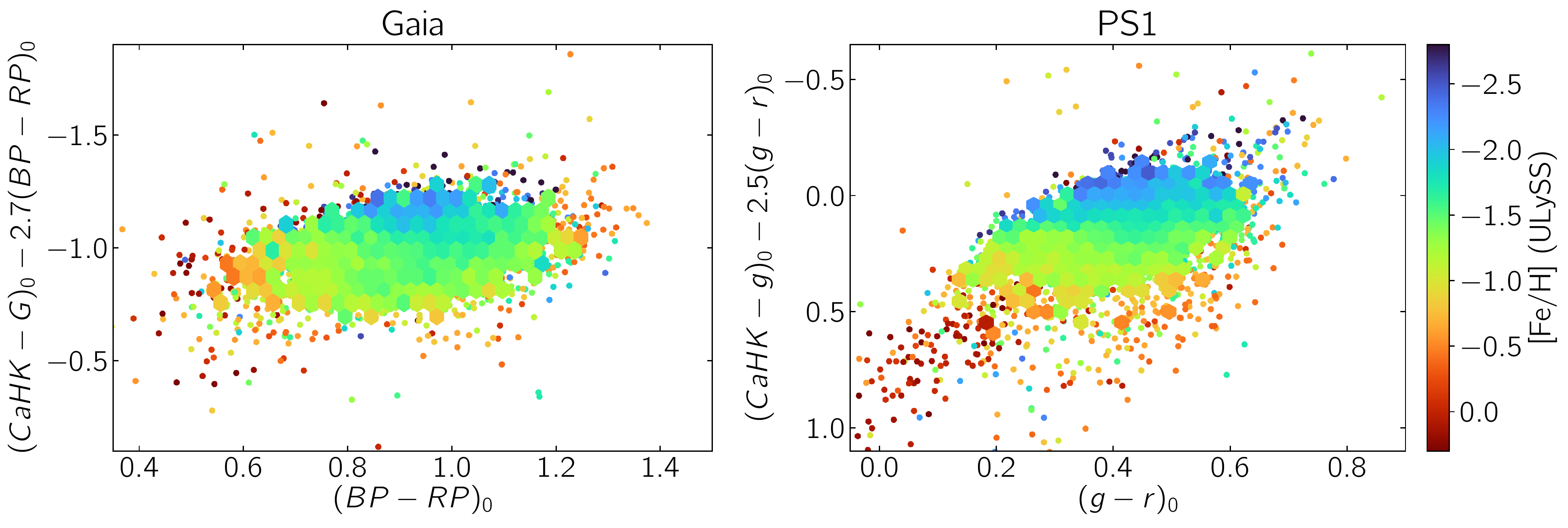}
\caption{Pristine colour-colour diagrams combining our $CaHK$ photometry with broad-band photometry from \Gaia (left) and PS1 (right). Top row: stars within our selected magnitude range and passing the parallax cut for a two-degree field of view centred on $(l,b)=(2.4^{\circ},-5.2^{\circ})$. We show our spectroscopic follow-up for this field, where the selection was originally made using \Gaia. These stars are colour-coded by their ULySS \feh (for details on the spectroscopic analysis, see Section~\ref{sec:uly}). The extinction vector for E(B$-$V) $= 0.45$ (the average in this field) is indicated by the arrow in the top left. Bottom row: 4300 stars in the Southern spectroscopic PIGS sample ($b < 0^{\circ}$) that pass the ULySS quality criteria. Each hexbin contains at least five stars and is coloured by its average ULySS \feh. Single stars are plotted as additional points.}
    \label{fig:selection}
\end{figure*}

\subsubsection{Calibration}

The (non-calibrated) catalogue produced after the data reduction steps described above needs to be calibrated with several additional steps. Firstly, because the MegaCam images show subtle variations in magnitude that depend on the location in the field of view,  a flat-field correction is applied as determined in Section~2.4.2 of S17b. Secondly, all MegaCam fields are put on the same scale. S17b used the determination of the stellar locus in the SDSS dereddened ($(g - i)_0, ($CaHK$ - g)_0$) colour-colour space to shift all fields to the scale of one reference field. The stellar locus consists mainly of metal-rich foreground stars. Unfortunately this method does not work for our footprint towards the inner Galaxy. The stellar locus is less well-defined since we look through a larger part of the disk, and, additionally, correcting for 3D extinction is a challenge. Thus, our MegaCam pointings are placed to create an overlap between neighbouring $CaHK$ fields of $\sim 0.1^\circ$, and these overlapping regions are used for a relative photometric calibration. All \Gaia DR2 source\_id's that have multiple $CaHK$ matches within 1'' are identified, and we assume that these sources are the same stars observed in two (or more) different MegaCam fields. These are then used to determine the relative offsets between fields. 

The 2017 and 2018 data sets and calibration are primarily used to select stars for our spectroscopic follow-up program (described in Section \ref{sec:spec}). As an initial (crude) calibration, we set one central field as the reference and started by determining offsets for the directly neighbouring fields with respect to the reference. For each pair of fields, all stars in common were used to compute the minimum $\chi^2$ and the best offset. This process was repeated for each of those fields and their neighbours that had not yet been calibrated, until all fields are on the same scale. In this calibration, each field is calibrated with respect to only one other field. 

In 2019, the calibration was changed to a more homogeneous approach, minimising the $\chi^2$ for the overlapping regions of all fields simultaneously (using only the stars with $CaHK$ uncertainties less than 0.01),
using the iterative method of \citet{magnier92} to find the best solution. We have since re-calibrated all our photometry using this approach, including 
the fields in the 2017 and 2018 footprint.

It is important to note that since we perform a relative calibration, the $CaHK$ photometry is not on the same scale as the $CaHK$ photometry in S17b.
It is only homogeneous for continuous regions; thus, fields with $b > 0^{\circ}$ and $b < 0^{\circ}$ are not necessarily on the same scale as the two regions do not overlap. 

\subsection{Target selection for spectroscopic follow-up}\label{sec:phottargets}

\subsubsection{Choice of broad-band photometry}

The calibrated $CaHK$ photometry was combined with broad-band photometry to determine the best metal-poor candidates. The $CaHK$ lines are not only sensitive to metallicity, but they also depend strongly on the temperature of the star, which can be approximated using broad-band colours. In S17b, SDSS photometry was used, but there is no SDSS photometry towards the Galactic bulge  \citep{sdss}. Instead, \Gaia DR2 \citep{gaia18} and Pan-STARRS1 \citep[PS1,][]{panstarrs} photometry were used to select targets. To select giant stars that are roughly at the distance of the bulge (for the typical extinction in our fields), we limited our magnitude range to $13.5 < G < 16.5$ when using \Gaia, and $14.0 < g < 17.0$ when using PS1 (see Section~\ref{sec:specstrategy}). An additional cut was imposed on the \Gaia parallaxes ($\varpi$) to remove foreground stars. This cut has slightly changed during the progress of the survey, from no cut in our spectroscopic follow-up pilot, in August 2017, to $\varpi > 0.25$~mas in 2018A, $\varpi -  \Delta \varpi > 0.30$~mas in 2018B and $\varpi -  \Delta \varpi > 0.25$~mas in 2019A.

The use of broad-band colour systems has evolved as well as we gained further insight by collecting larger spectroscopic samples. In 2018, after the pilot program, we enthusiastically switched from PS1 photometry to Gaia DR2 photometry when it became available, because the latter has the advantage of being more precise and having all-sky coverage. The bulk of our spectroscopic follow-up to date has thus been selected with the help of \Gaia photometry. Further analysis of the results of these \Gaia-selected runs, however, showed that the PS1 photometry performs better for our purposes, especially in regions with high extinction. The reason for this is that even though the \Gaia photometry is much more precise than PS1, it is much harder to correct for extinction. We have used \Gaia extinction coefficients determined by F. Anders
\footnote{\label{extnote}These coefficients are $A_G/A_V = 0.718$, $A_{BP}/A_V = 0.9785$ and $A_{RP}/A_V = 0.576$. They were derived by interpolating between coefficients for stars of 4000~K and 6500~K. The \citet{schlafly11} value for $A_V = 2.742$ ($R_V = 3.1$) was used.} for stars of \teff = 5250~K and an extinction of $A_V$ = 1.5 (which corresponds to an E(B$-$V) $\approx 0.5$), together with the \citet{green18} reddening map. Because the \Gaia filters are so broad, their extinction coefficients depend non-negligibly on the amount of extinction itself and on the temperature of the star. The coefficients are therefore not constant throughout our footprint, which covers a range of extinction values.

For these reasons, we switched back to using PS1 for our follow-up in 2019 and indeed noticed that it reduced the contamination from more metal-rich stars. The PS1 filters are narrower and therefore less dependent on the temperature of the star and the extinction. Additionally, the \citet{green18} reddening map has been made using PS1 photometry and they provide up-to-date PS1 extinction coefficients to go with this map, making the broad-band de-reddened photometry more self-consistent. For the $CaHK$ photometry, the extinction coefficient 3.924 from S17b was used, which is on the \citet[][hereafter SF11]{schlafly11} scale. 

\citet{green18} provide a 3D map that can be queried at different distances. For our purposes, we assume that all dust is located in the foreground disk (which should be valid in the PIGS latitude range), and query the map at a distance of 8~kpc. Another assumption is that the extinction law is constant (and correct). However, \citet{schlafly16} have shown that there appear to be spatial variations, which have not yet been characterised in detail. Finally, the extinction correction is limited by the pixel scale of the map used, while the extinction may vary on even smaller scales.

In this paper, whenever photometry is shown (in Figure~\ref{fig:selection} and \ref{fig:selectioneff2}), the updated PS1 reddening map by \citet[][hereafter G19]{green19} was used. The difference between the extinction coefficients for the PS1 $g$ and $r$ filters provided by the authors and those from SF11 is significant (0.35 for both), due to differences in the adopted reddening vector and scaling of their map. The extinction coefficients for $CaHK$ and the \Gaia photometry are, however, on the SF11 scale. 
To convert the \Gaia photometry to the G19 reddening scale, 0.35 was added to the $A_V$ as used in Footnote~\ref{extnote} (assuming that the difference is same as for the PS1 $g$ and $r$ filters, which overlap in wavelength range with $V$).
The $CaHK$ wavelength range lies outside the PS1 coverage and is bluer (more affected by extinction), therefore extrapolation is likely invalid for this filter. To estimate the G19 $CaHK$ coefficient, we instead tested a range of coefficients and determined for which value the dispersion in our photometric metallicity bins was the smallest (see Section~\ref{sec:photeff} for a description of the photometric metallicities). A minimum was found for $A_{CaHK} = 4.5$, which was therefore adopted as the coefficient to use with the G19 map.

\subsubsection{Metallicity-sensitive photometric selection}

In S17b, photometric metallicities for individual stars were determined from the $CaHK$ and broad-band photometry. This was not done for PIGS for multiple reasons: our $CaHK$ photometry is not all on the same absolute scale, the method had not been calibrated for \Gaia broad-band photometry, and the photometric metallicities would in any case be much less precise than in the halo because of uncertainties in the reddening correction. 

Instead, a relative selection was applied in each follow-up field. An example selection for follow-up with \Gaia is shown in the top left panel of Figure~\ref{fig:selection}, for a two-degree field of view. In this Pristine colour-colour diagram, the metal-poor stars are roughly expected to lie towards the top. Targets were therefore simply selected by their vertical location in this diagram, within a certain $(BP - RP)_0$ range corresponding to the expected location of metal-poor giant stars. The coloured points are stars followed up spectroscopically, which have been colour-coded according to their spectroscopic metallicity (for details on the spectroscopy, see the next section). The top-right panel of the figure shows the same stars, but now with PS1 broad-band photometry instead of \Gaia. The contamination of metal-rich stars would have been lower, if this field had been selected using PS1. In the bottom row of the figure we present the same diagrams, now for all stars in our spectroscopic follow-up program with $b < 0^{\circ}$ (a homogeneously calibrated region of the photometry). There is a much clearer gradient with \feh in the case of PS1 photometry compared to the \Gaia photometry. The main culprit of this difference is the extinction, which -- as explained above -- is difficult to correct for in the broad-band \Gaia filters. 

\section{Spectroscopic follow-up observations}\label{sec:spec}

We performed low-/intermediate-resolution spectroscopic follow-up of our metal-poor candidates with the 3.9m Anglo-Australian Telescope (AAT). The follow-up was done with the dual-beam AAOmega spectrograph \citep{saunders04} with the Two Degree Field (2dF) multi-object instrument, which has roughly 360 fibres for science targets (plus $\sim$40 sky and guide fibres) over a 2-degree diameter field of view \citep{lewis02, sharp06}. To date, 23 fields have been observed, yielding spectra for 8350 stars. The right panel of Figure~\ref{fig:footprint} shows the current spectroscopic PIGS footprint. Three fields were observed in service mode in August 2017 and June 2018, the rest was observed during three observing runs in June 2018, August 2018 and April 2019. The observations of two fields in August 2017 were our spectroscopic pilot program. 
 
In this section, our observing strategy, the data reduction, the determination of radial velocities, and two independent methods of analysing the spectra are described. We first describe the derivation of \teff, \logg and \feh with ULySS \citep{koleva09}, which is a data-driven method using empirical reference model spectra. In our second analysis, the employed method is FERRE \citep{allende06} with a grid of synthetic model spectra to determine \teff, \logg, \feh, and \cfe. The parameters from ULySS for the sample in this work were used in \citet{arentsen20}, although we note that if the FERRE parameters had been used instead this would not have altered the results of that paper.
 
\subsection{Observing strategy}\label{sec:specstrategy}

For each field, the top 200 candidates from the photometry were provided and they were given the highest priority in the configuration of 2dF. Generally, $\sim$95\% of those candidates were allocated a fibre. The next best 200 candidates were provided with intermediate priority and the next best 200 candidates with low priority to fill the remaining $\sim 160$ science fibres. Stars already in the EMBLA survey were removed from our selection in most fields. 25 fibres were allocated to sky positions, the input catalogue for these was created by hand using the Aladin tool \citep{aladin1} and PS1 images. Finally there are 8 fibres for guide stars, which were selected from \Gaia DR2. 

The stars were observed with the 580V and 1700D gratings, resulting in blue spectra at a resolving power R$\,\sim\,$1300 covering $\,\sim\, 3700 - 5500$\,\AA, and near-infrared calcium triplet (CaT) spectra at R$\,\sim\,$11\,000 covering $\,\sim\, 8400 - 8800$\,\AA. The CaT spectra were used to determine radial velocities, and both spectra together to determine stellar parameters and metallicities. 

\begin{figure}
\centering
\includegraphics[width=0.8\hsize,trim={0.0cm 0.0cm 0.0cm 0.0cm}]{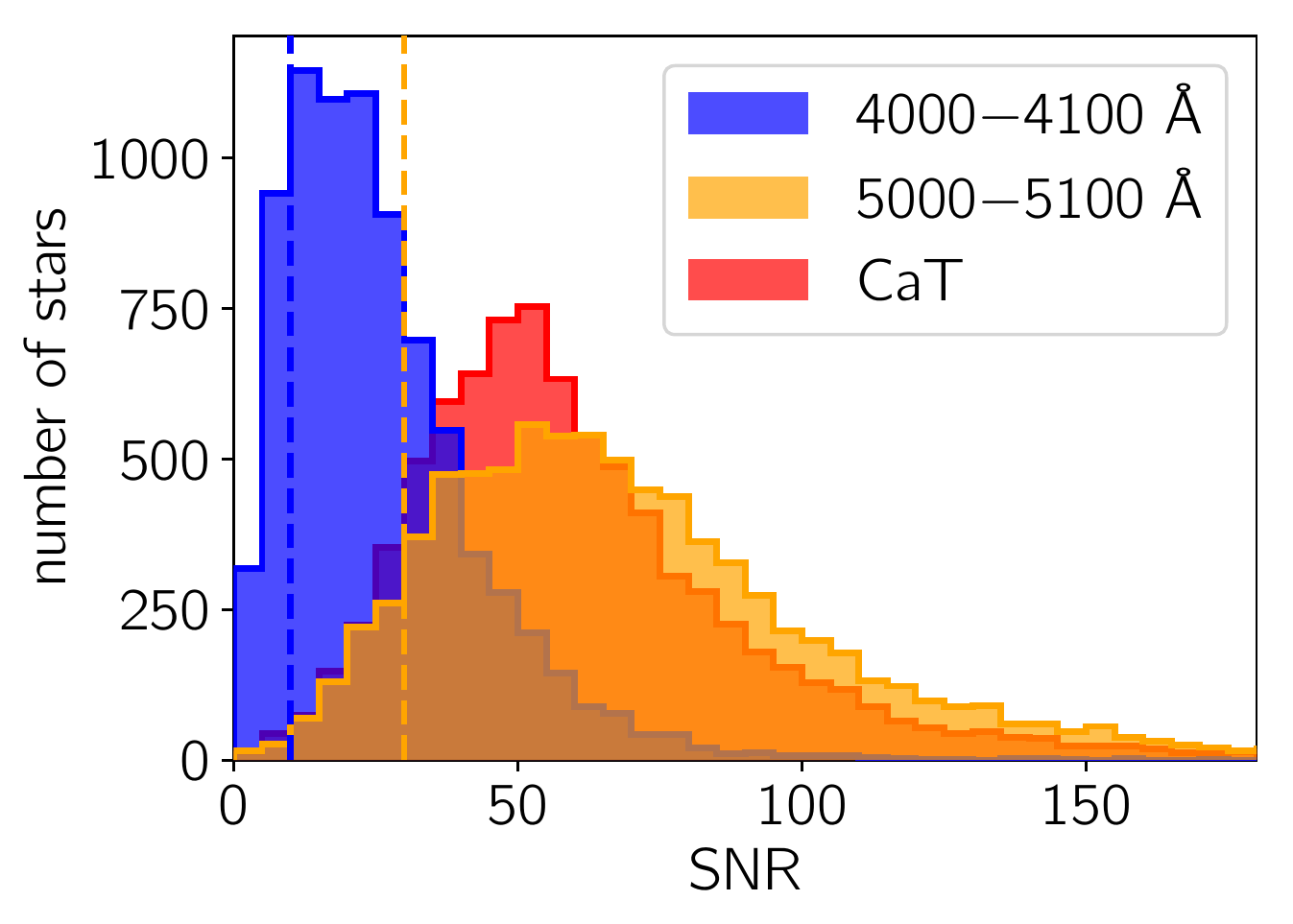}
\caption{Distribution of the SNR in different regions of the AAT spectra. The CaT range is from $\sim 8400-8800$~\AA. The SNR$_{4000-4100} = 10$ and SNR$_{5000-5100} = 30$ quality cuts (see Section~\ref{sec:uly}) are indicated by the vertical dashed lines.}
    \label{fig:SNR}
\end{figure}

The recommended magnitude range for stars in one 2dF field is $\sim$3 magnitudes. When using \Gaia photometry, stars between G~=~13.5 and 16.5 were selected, and when using PS1, the selected stars were between g~=~14.0 and 17.0. We aimed for a minimum signal-to-noise ratio (SNR) of 20 in the blue spectra, which meant integrating for 2 hours on each field. Because there were some fibre throughput issues (see below), this was not always successful for the faintest stars. The SNR in different regions of the spectra is summarised in Figure~\ref{fig:SNR}.

\subsection{Data reduction}\label{sec:specreduction}
The spectra were reduced using the AAT \textsc{2dfdr}\footnote{\url{https://aat.anu.edu.au/science/software/2dfdr}} (v. 6.46) package. The standard settings for the 580V and 1700D spectra were used respectively, with the exception of two parameters that determine how spectra from multiple observations are combined. 

The first setting which was changed is the FLUX\_WEIGHT parameter, which determines how much weight is given to each individual exposure when combining to a single spectrum. 
The standard setting determines a frame zero-point by taking the median of the zero-points for each fibre. These fibre zero-points are derived using the difference between an empirical brightness from a smoothed version of the spectra and the expected brightness from provided external photometry. The frame zero-points are used to determine the relative weights of multiple exposures. 
This, however, is inappropriate when there is variation in throughput across the field. From the \textsc{2dfdr} quality plots, we concluded such a variation is indeed present in most of our fields. There are often regions on the plate that have a reduced throughput. Although these regions are not always in the same location, they are usually in the outer regions of the plate. The effect is described in some detail in the appendix of \citet{Li19}. The root cause of the problem is the accuracy of the fibre placement, but exactly why the fibre placement is bad is unknown.
There have been efforts at the AAT to resolve these problems, but so far no solution has been found (priv. comm. T. S. Li and C. Lidman). 
Because of the observed variation in throughput, we therefore decided to use the flux weighting by object instead of by frame, where each object is assigned its own weight in the combination of multiple frames. 

Secondly, the ADJUST\_CONTINUUM option was turned off. It was originally introduced into the software to fix a problem with the continuum of early 2dF spectra, which was solved in 1999. The \textsc{2dfdr} documentation suggests it is worth using for data taken after the problem was fixed (it is the default), but it can also be turned off. We found that the default setting sometimes produced spurious, unphysical calcium triplet line shapes, therefore this option was disabled. 

\subsection{Radial velocities}\label{sec:specrv}

Radial velocities were determined from the CaT spectra using the \textsc{fxcor} package in \textsc{IRAF}\footnote{IRAF (Image Reduction and Analysis Facility) is distributed by the National Optical Astronomy Observatories, which are operated by the Association of Universities for Research in Astronomy, Inc., under contract with the National Science Foundation.}. The CaT spectra were cross-correlated with synthetic spectra, which were created using the MARCS~(Model Atmospheres in Radiative and Convective Scheme) stellar atmospheres and the Turbospectrum spectral synthesis code \citep{alvarez98,Gustafsson08,Plez08}. A preliminary velocity was derived using a template with $\teff = 5000$~K, $\logg = 2.5$ and $\feh = -2.0$. After zero-shifting the spectra, we derive stellar parameters with ULySS (see Section~\ref{sec:uly}). To improve the radial velocities, they were re-derived using the template that is the closest match to the ULySS parameters, from the following grid: $\teff = [4500,5000,5500]$ K, $\logg = 2.5$ and $\feh = [0.0,-1.0,-2.0,-3.0]$. From the \textsc{fxcor} formal uncertainties and a test of repeated observations, the radial velocity uncertainties are estimated to be of the order of 2~\kms.

\subsection{Spectral analysis with ULySS}\label{sec:uly}

\subsubsection{Method}

For the first determination of stellar atmospheric parameters (\teff, \logg and \feh) the University of Lyon Spectroscopic analysis Software (ULySS\footnote{ULySS is available at \url{http://ulyss.univ-lyon1.fr}}, \citealt{koleva09}, version 1.3.1) was used. ULySS is a full-spectrum fitting package that determines the best parameters by minimising the $\chi^2$ of the residuals of the fit between an observed spectrum and model spectra. The model library used is the empirical MILES library \citep{sanchez06, falconbarroso11} with a resolving power of $\sim$2200, for which \citet{prugniel11} have computed a ULySS interpolator. We use the version of the interpolator as updated by \citet{sharma16}, who improved it for cool stars. Recently, \citet{arentsen19_xsl} have used ULySS with this interpolator to determine stellar atmospheric parameters for DR2 of the X-shooter Spectral Library \citep{gonneau20}.
The reliability of the method has been tested extensively in the aforementioned works, and it has been found to be robust for FGK stars with $-2.5 < \feh < 0.5$ and $0 < \logg < 5$. Details on how the method works can be found in \citet{prugniel11}, it is briefly summarised here. 

The model spectrum $S(\lambda)$ in ULySS is described by:
\begin{equation}
    S(\lambda) = P_n(\lambda) \times G(v_{\mathrm{r}}, \sigma) \otimes \mathrm{TGM}(\teff, \logg, \feh,\lambda),
\end{equation} 

\noindent where $P_n(\lambda)$ is a series of Legendre polynomials up to degree $n$, and $G(v_{\mathrm{r}}, \sigma)$ is a Gaussian broadening function described by the radial velocity $v_{\mathrm{r}}$ and the broadening width $\sigma$. The $\sigma$ is mainly related to the resolution of the spectra and to the rotational velocity of a star, sometimes it is also increased to smooth the spectra if no good fit can be found otherwise. 
The polynomial $P_n(\lambda)$ normalises the spectra, correcting for, e.g., the effects of the detector response and dust extinction. 
The TGM function represents the interpolator, which is a polynomial function that creates spectra from a reference library for a given combination of \teff, \logg and \feh. 
In the fit, the free parameters are \teff, \logg, \feh, $v_{\mathrm{r}}$, $\sigma$ and the coefficients of $P_n(\lambda)$. 

The spectra reduced to the rest-frame with our preliminary CaT radial velocities were used, therefore $v_{\mathrm{r}}$ only absorbs uncertainties in the velocity. The degree of $P_n(\lambda)$ was set to 10, which is enough to correct for the response function of the detector and the extinction, but not so high that it starts affecting the parameters. The spectra were fit between $3800-5400$~\AA. The region of the CH G-band was masked between 4260 and 4325 \AA, since carbon can be anomalous in metal-poor stars and it should be avoided that this drives the fit. The MILES library does not contain the region of the CaT, so the blue arm was fit alone.

We run ULySS with a grid of initial guesses to avoid getting trapped in local minima. The starting grid is $\teff = [4000,5000,6000]$~K, $\logg = [2.0,4.0]$, and $\feh = [-2.4, -1.7, -0.3]$, and the solution with the smallest $\chi^2$ is finally adopted as the best solution. Furthermore, the \texttt{/CLEAN} option was turned on, which rejects regions where the fit is bad due, for example, to bad pixels or features which are not in the model spectra such as strong carbon bands.

Some spectra do not pass through this first run or have extremely high $\sigma$ values. They were rerun with a multiplicative polynomial of degree $n=30$, which gives solutions for 89 additional stars. The main reason that these spectra fail, is that they either have a strong continuum shape that cannot be corrected with $n=10$ or they are very carbon rich, in which case the higher polynomial degree tries to correct for the carbon features.

\begin{figure*}
\centering
\includegraphics[width=0.93\hsize,trim={0.0cm 0.0cm 0.0cm 0.0cm}]{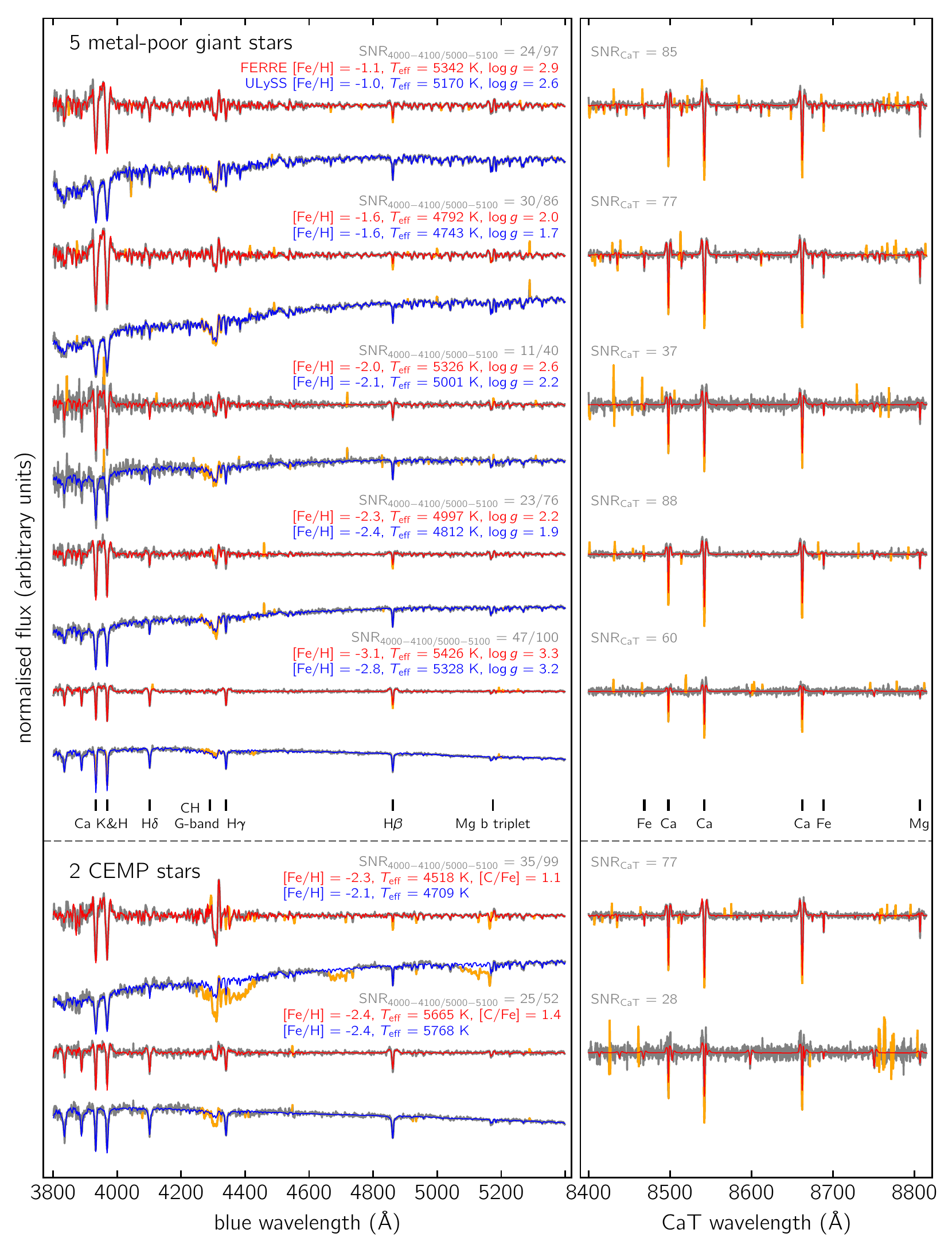}
\caption{Example fits for ULySS (blue) and FERRE (red) for five metal-poor giants and two CEMP stars. For both analyses, the observed (normalised) spectrum is shown in grey. Regions in the observed spectra which are masked in the fit are indicated in orange, these contain both a priori masks (the CH G-band in ULySS and the calcium triplet line cores in FERRE) and pixels that are masked iteratively. Several of the strongest spectral features have been indicated. For each star, the stellar parameters from both analyses are shown.}
    \label{fig:fits}
\end{figure*}

Example fits for seven stars are presented in Figure~\ref{fig:fits}, in blue. 

\subsubsection{Selection of reliable results}

The atmospheric parameters were determined for 8089 PIGS stars. To select reliable solutions, the following criteria were used: 

\begin{itemize}
\item SNR$_{4000-4100}$ > 10 or SNR$_{5000-5100}$ > 30
\item signal-to-residual ratio (SRR) > 15
\item $\sigma$ < 150 \kms 
\item fraction of rejected pixels in the fit between $4600-4800$~\AA $< 10 \%$ and between $5000-5200$~\AA also $< 10\%$
\item not double-lined
\end{itemize}

\noindent which leaves 7126 stars. The SNR cut was determined using repeated observations:  above these SNR values, the dispersion in the results for individual observations is almost independent of SNR. This cut removes 677 stars, a relatively large fraction because of the flux-loss issues described in Sect.~\ref{sec:specreduction}. The second and third cuts were set after inspecting the fits of stars that passed the SNR criterion, but had low SRR or very high $\sigma$ which typically indicates a bad fit. Then the fraction of rejected pixels in the fit in the regions of two strong carbon bands were used to weed out a number of carbon-enhanced stars that had not been removed yet by the other cuts. The ULySS results are not reliable for such stars. These three cuts remove an additional 136 stars. The final cut removes stars with double-lined CaT spectra, based on a cross-correlation function (CCF) analysis. We cross-correlate the observed spectrum and the radial velocity synthetic template, and flag stars that have two or more clear peaks in the CCF (if the secondary is at least 50\% as strong as the primary). This removes a final 150 stars. Given the crowding towards the bulge, the double-lined stars are probably more likely to be chance-alignments than spectroscopic binaries.
 
\subsubsection{Stellar parameter uncertainties}\label{sec:ulyssunc}
 
The formal uncertainties provided by ULySS might under- or overestimate the statistical uncertainties if the uncertainty spectra are not perfect. To investigate this, a repeat- observation test as in \citet{arentsen19_xsl} was used. For each independent pair of observations of the same star, we computed the following weighted difference in the stellar parameters:

\begin{equation}
    \Delta \mathrm{P}_{w,i} = \frac{\mathrm{P}_{1,i} - \mathrm{P}_{2,i}}{\sqrt{\epsilon_{1,i}^2 + \epsilon_{2,i}^2}},
    \label{eqn:DP}
\end{equation} 

\noindent where P represents \teff, \logg or \feh, $\epsilon$ represents the formal uncertainty given by ULySS, and 1 and 2 indicate two observations of the same star. The distribution of $\Delta \mathrm{P}$ should be close to Gaussian, with a standard deviation $\sigma_{\Delta \mathrm{P}} = 1$ if the uncertainties are properly scaled. A smaller/larger $\sigma_{\Delta \mathrm{P}}$ indicates over-/underestimated uncertainties. 

We select two fields, one with four sub-exposures (F\_273.7-27.1) and one with six sub-exposures (F\_284.0\_30.0). The stars in these fields are a representative sub-sample of the total. After applying the quality cuts for reliable solutions from the previous section, 2850 independent pairs of observations remain, which were used for the test. For all three parameters the distribution of $\Delta \mathrm{P}$ is found to be close to Gaussian, with a standard deviation of 1.2. This indicates that the uncertainties are somewhat underestimated, therefore we multiply the formal ULySS uncertainties by this factor to derive proper internal uncertainties. The median of the internal uncertainties for the reliable set of ULySS parameters are 32~K, 0.08~dex and 0.04~dex for \teff, \logg and \feh, respectively.

\subsection{Spectral analysis with FERRE}\label{sec:ferre}
\subsubsection{Method}

The spectra were additionally analysed with the FERRE\footnote{FERRE is available from \url{http://github.com/callendeprieto/ferre}} code \citep{allende06}. FERRE performs full-spectrum fitting against a synthetic stellar model library, interpolating between the different nodes (the cubic algorithm was selected for the interpolation). In addition, the Boender-Timmer-Rinnoy-Kan (BTRK) algorithm was used as the global optimisation technique already included in the FERRE architecture. For each fit four different runs have been performed to avoid getting trapped in a local minimum. The \teff, \logg, \feh and \cfe were derived simultaneously. In this work, we mainly focus on \teff, \logg, and \feh. The results with regards to \cfe merit a separate discussion that is the subject of a forthcoming paper (Arentsen et al., in prep.). The synthetic spectra in the reference grid were computed with the ASSET code \citep{koesterke08}, with the model atmospheres computed using the Kurucz code, described in \citet{meszaros12}, assuming a microturbulence of 2~\kms. The grid is similar to the carbon-rich ultra metal-poor (CRUMP) grid presented in \citet{aguado17}, which has also been used in previous \textit{Pristine} analyses \citep{youakim17, aguado19}, with two important differences: our current grid starts at $\teff = 4500$~K instead of $4750$~K, and it extends up to $\feh=+0.5$ compared to $-2.0$ previously. The grid ranges and grid steps are the following:

\begin{itemize}
\item $4500~\mathrm{K} < \teff < 7000~\mathrm{K}$, $\Delta \teff = 250~\mathrm{K}$
\item $1.0  < \logg < 5.0$, $\Delta \logg = 0.5$
\item $-4.0  < \feh < +0.5$, $\Delta \feh = 0.5$
\item $-1.0  < \cfe < +3.0$, $\Delta \cfe = 1.0$
\end{itemize}

\noindent where for all nodes $\alphafe = +0.4$. This value of \alphafe is reasonable for bulge stars with $\feh < -1.0$ or even $-0.5$ \citep{barbuy18}, but may result in biased results for more metal-rich stars. However, as the focus of this work is on metal-poor stars, this grid is the best choice for our science case. 
One of the advantages of the FERRE code is that it is able to fit different regions of the spectrum simultaneously even if they have different resolutions. That fact allows us to use all the information contained in both arms, fitting the blue and the CaT spectra together. We found, for instance, that including the CaT spectra reduces the scatter in the determined \logg values considerably, likely because the wings of the calcium lines are sensitive to gravity.
The blue and CaT spectra were corrected to rest-frame velocities using the CaT radial velocities. Several artefacts were removed from the spectra which are the result of dead pixels, and usually occur in the same place in the spectrum. The model spectra were smoothed to $R=1300$ and $R=11\,000$ for the blue and CaT spectra, respectively. The observed and model spectra were normalised using a running mean filter of 30 pixels.  

We ran FERRE twice. First FERRE is run to get an estimate of the stellar parameters and a corresponding model spectrum for each observed spectrum. The cores of the calcium triplet lines were masked because they are strongly affected by non-local thermodynamical equilibrium (NLTE) processes\footnote{The line wings beyond 1.5~\AA from the line centre can be safely used in LTE analyses, therefore the central 1.5~\AA of each line were masked. However, because of the running mean normalisation, some of the NLTE information from the core may be spread to the neighbouring pixels. To test whether this significantly affects the analysis, we compare the results from FERRE runs with the blue arm only and with both arms together. There are no significant systematic changes in the metallicities between the two runs, although there are some systematic differences in \logg. It is unclear whether this is because the CaT spectra add necessary information to get better \logg values, or whether this is a NLTE effect. Since there are no significant changes in the metallicities, we assume that it is sufficient to mask the cores of the lines.}. 
The residuals between the observed and model spectrum were used to flag bad pixels. In the blue spectra, the standard deviation $\sigma_f$ between model and observed spectrum was determined in chunks of $200$~\AA, because the SNR varies strongly throughout. Pixels were flagged as bad if they are discrepant by more than $3\sigma_f$, and added their neighbouring pixels if they are discrepant by more than $2\sigma_f$. Flagged regions are typically dead pixels that had not yet been removed, data reduction artefacts or other outliers. For the near-infrared spectra a single $\sigma_f$ was computed since the SNR does not vary, and the pixels were flagged in the same way. The threshold was increased to $5\sigma_f$ around the three calcium triplet lines and the magnesium line at $8806$~\AA, to avoid these lines being removed entirely. 
The SNR is generally high enough that uncertainties in the spectral modelling start to affect the fit of the lines. Typically more pixels are flagged in the CaT spectra compared to the blue spectra because of faulty sky subtraction, which affects the infrared more. When we run FERRE for the second time, the uncertainties on the pixels flagged as bad are increased to $10^9$, in order to set their weight in the fit to practically zero. This second run resulted in significantly better fits for many stars. 
Example fits for seven stars are presented in Figure~\ref{fig:fits}, in red. 

\subsubsection{Selection of reliable results}
Reliable stellar parameters were selected using the following criteria:

\begin{itemize}
\item SNR$_{4000-4100}$ > 10 or SNR$_{5000-5100}$ > 30
\item $\log \chi^2 < -2.4$
\item not double-lined
\end{itemize}

\noindent which leaves 7186 stars. The SNR cut is the same as for the ULySS analysis, as is the cut on double-lined CaT spectra. The cut on the $\log \chi^2$ of the fit has been set to remove stars that are outliers by $>+3\sigma$ from the median. It removed 60 stars on top of the two other cuts. 

\subsubsection{Stellar parameter uncertainties}\label{sec:ferreunc}

The same method was used as with the ULySS analysis to investigate the magnitude of the formal FERRE uncertainties, using 2785 independent pairs of observations. We find close-to-Gaussian distributions of $\Delta \mathrm{P}$ (see Equation~\ref{eqn:DP}) with a standard deviation of 0.1 for each of the three main parameters (\teff, \logg and \feh) and 0.07 for \cfe. This means that the formal FERRE uncertainties are severely overestimated, by a factor of $\sim$10. No satisfactory explanation for this has been found. After re-scaling the formal uncertainties into internal uncertainties using the factors derived from the repeated-observations test, the values appear reasonable. The median internal FERRE uncertainties are 105~K, 0.34~dex, 0.09~dex and 0.12~dex for \teff, \logg, \feh and \cfe, respectively. This is larger than for the ULySS analysis. 

\begin{figure*}
\centering
\includegraphics[width=0.49\hsize,trim={0.0cm 0.0cm 0.0cm 0.0cm}]{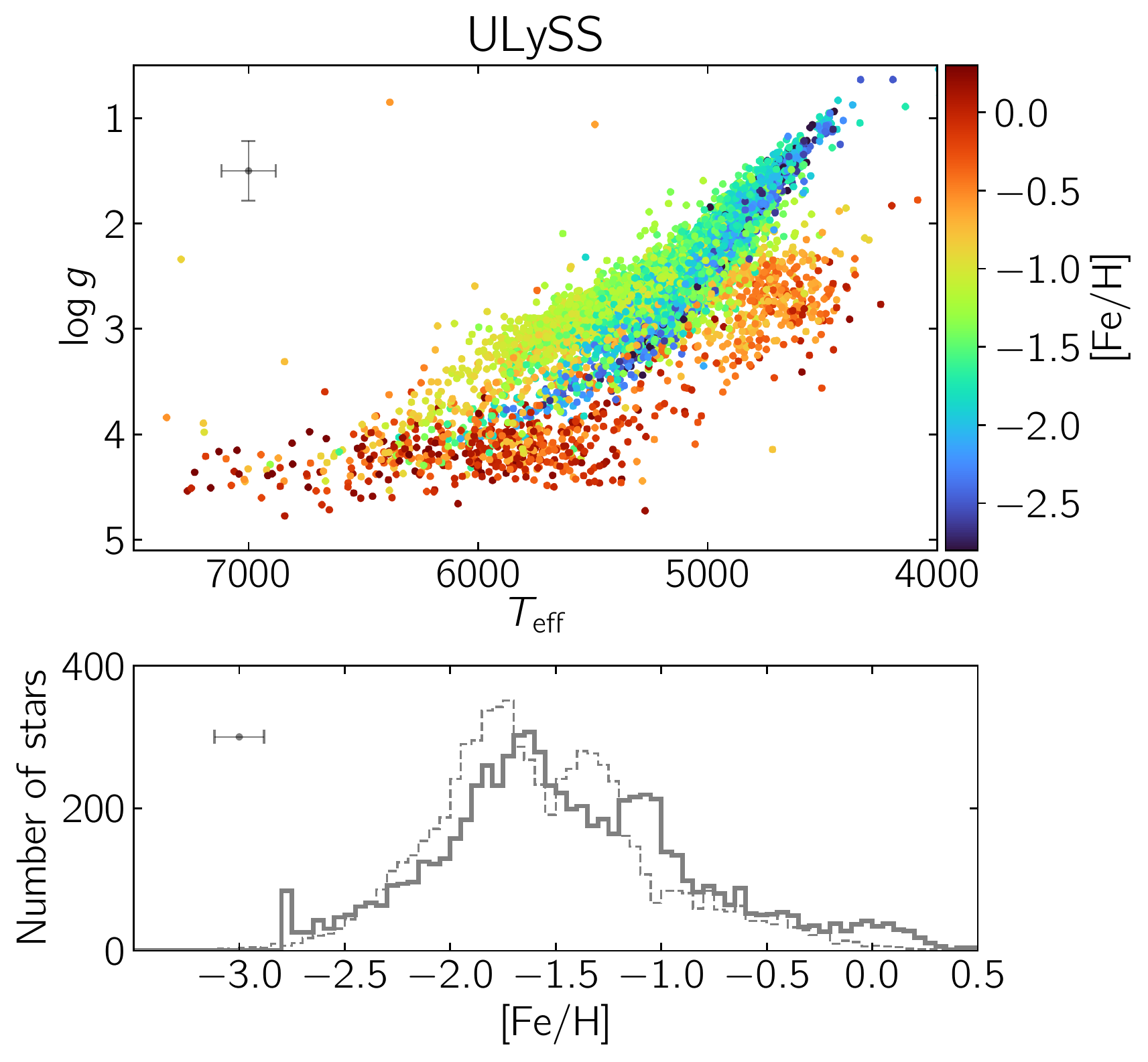}
\includegraphics[width=0.49\hsize,trim={0.0cm 0.0cm 0.0cm 0.0cm}]{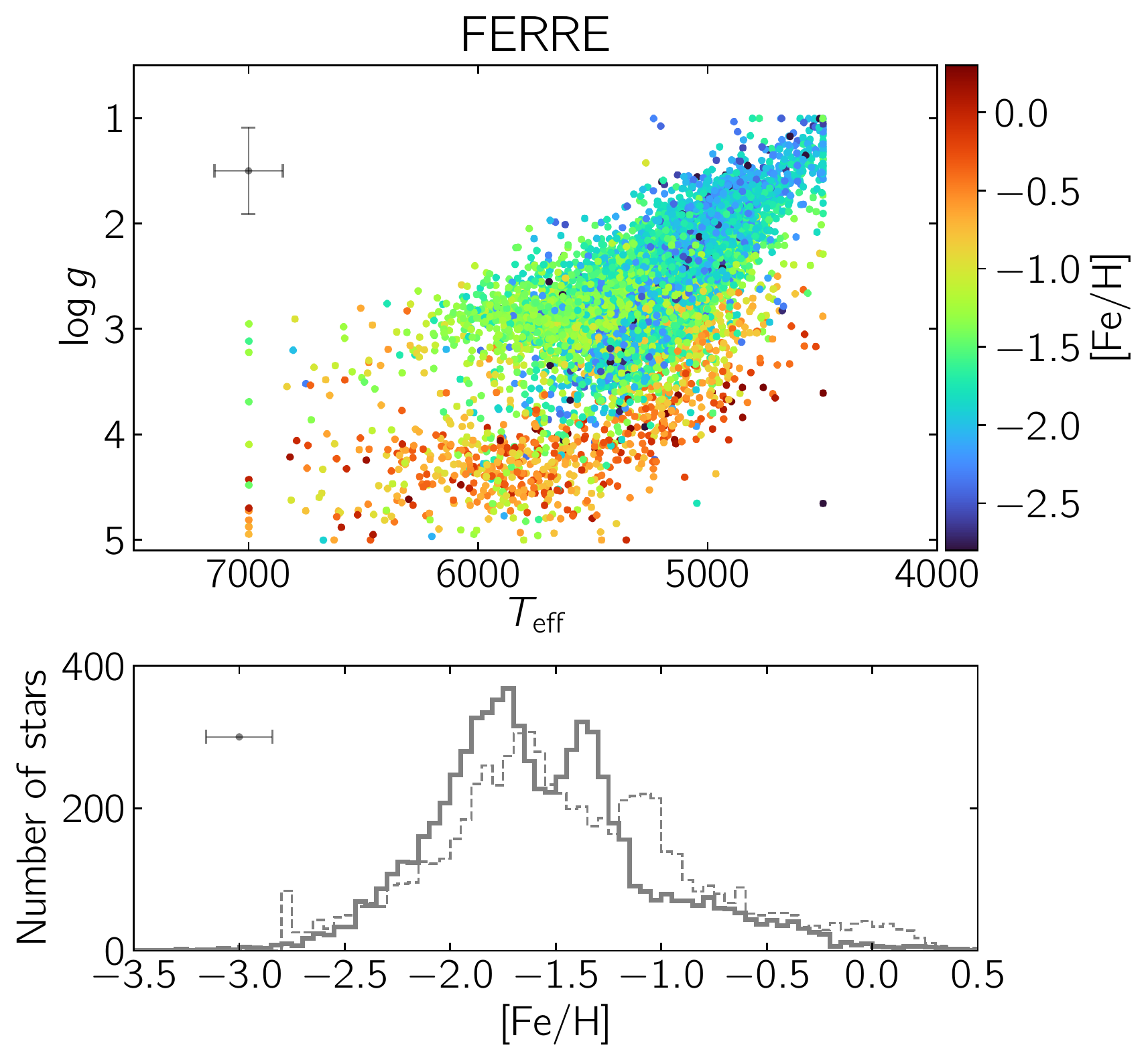}
\caption{Top row: Kiel diagram, colour-coded by \feh from ULySS (left) and FERRE (right). Each panel shows the stars that pass the quality criteria for both methods. Bottom row: metallicity distributions, where dashed histogram indicates the distribution with the other method. Median final uncertainties for metal-poor ($\feh < -1.0$) giants are shown in the top left corner of each panel. There are three grid effects visible in this figure: the \feh limit in the ULySS analysis is $-2.8$, and the FERRE grid limits are $4500 < \teff < 7000$~K and $1 < \logg < 5$.}
    \label{fig:HR_UF}
\end{figure*}

\section{Stellar parameter results}\label{sec:speccomp}

The results of the ULySS and FERRE analyses of the PIGS low-/intermediate-resolution spectra are presented here. They are compared to each other to identify regions where the methods have issues and to derive the total uncertainties. The results are also compared to those in the SDSS APOGEE database for a sub-sample of stars. 

\subsection{Metallicity distribution and Kiel diagram}\label{sec:HR}

There are 7073 PIGS stars that passed the quality criteria for both the ULySS and FERRE analysis methods. The Kiel diagrams from ULySS and FERRE for these stars are shown in the top row of Figure~\ref{fig:HR_UF}, the metallicity distributions in the bottom row. There are several noteworthy features in the Kiel diagrams and metallicity distributions, seen in both analyses. There are a number of metal-rich dwarf stars ($\feh > -0.5$, $\logg > 3.5$) in our sample, which are in the sample mainly due to the \Gaia selection (as can be seen in Figure~\ref{fig:selection}; and more of these entered our selection box in higher extinction regions), or due to the selection testing in our two pilot fields. However, most of the stars in the sample are giants, consisting of three types. Firstly, there are the metal-poor red giant branch (RGB) stars which were the target of our survey ($\feh < -1.5$). These are the most numerous, and correspond to the largest peak in the metallicity distributions in Figure~\ref{fig:HR_UF}. Secondly, there are the hotter horizontal branch (HB) stars, with slightly higher metallicities ($ -1.5 < \feh < -1.0$). These are unintentionally part of our selection because they are relatively blue in colour and therefore also enter our selection box. They are less numerous than the RGB stars, but constitute a significant fraction of the sample and make up the second peak of the metallicity distribution. These stars are extremely useful, since they are metal-poor standard candles. Finally, there is a small number of metal-rich giants ($\feh > -1.0$), the result of ``normal'' contamination in our photometric selection. There is such a large number of metal-rich stars in the bulge that it is to be expected that some of them scatter into our selection box by accident, especially in higher extinction regions. 

There are clear differences between the ULySS and FERRE analyses, which are the result of different strengths and limitations of the two methods. The features in the ULySS Kiel diagram are narrower, which we attribute to the data-driven nature of this method. Data-driven methods have two properties that contribute to creating such narrow features: the reference library contains only ``physical'' stars that do indeed exist, and the differences in stellar parameters between each of the library spectra is irregular, i.e., not constant. The latter aids in the quality of the results by learning to interpolate between smaller offsets, unlike the use of synthetic grids, which have regular grid step spacing, thereby missing non-linear interpolation opportunities.  Also, synthetic grids cover all possible combinations of 
stellar parameters (even those that should not exist), providing a larger parameter space for exploration, even non-physical regimes.  And finally, most synthetic spectral grids are missing relevant physics, such as non-LTE corrections for some elements which can affect the stellar parameters \citep[e.g.,][]{kovalev19}.

Significant efforts are underway to create and/or extend existing empirical stellar spectral libraries to include more of the rarer stars \citep[e.g. XSL,][]{gonneau20}, although they remain limited. In the PIGS sample, this limitation is particularly relevant for the HB stars and for the most metal-poor stars. In the MILES library and its interpolator (which is used in the ULySS analysis), only a limited number of HB stars are present, all with similar metallicities \citep[$\feh \approx -1.1$,][]{prugniel11}. It is therefore suspicious that in the ULySS analysis, we find a peak of HB stars at exactly the same metallicity. The FERRE analysis should be unbiased in this regime, and finds the HB metallicities to be $\sim 0.3$~dex lower. There are also only a handful of very metal-poor stars ($\feh < -2.5$) in MILES. The limit of the interpolator is at $\feh = -2.8$, hence the spike in the ULySS metallicity distribution there. In contrast, the FERRE synthetic grid goes to $\feh = -4.0$, therefore the results for the metal-poor tail are expected to be more trustworthy.

We highlight three final differences between the ULySS and FERRE analyses. First, the metal-rich dwarfs are more metal-rich in ULySS than in FERRE. This is due to the limitations of the FERRE grid chosen for this analysis, which only contains $\alpha$-enhanced spectra. In our selected magnitude range, dwarf stars lie between us and the bulge and are likely part of the disk, and metal-rich disk stars are typically not enhanced in $\alpha$-elements. Therefore our FERRE analysis does not derive the proper metallicity for these stars. Second, there are a number of relatively metal-rich stars that ULySS classifies as giants with $\logg \sim 3.0$, and FERRE as dwarfs/sub-giants ($\logg > 3.5$). Inspection of the spectra of these stars shows that they have Ca H\&K in emission: these are active stars. This also explains why they ended up as contamination in our photometric selection: the emission filled in the Ca II H\&K absorption lines, making them look metal-poor in the $CaHK$ photometry. Retrospectively, we found that many of these stars can be removed using a \Gaia DR2 selection cut based on a combination of the mean flux over error for the $G$-band and the number of repeated observations. Finally, there are two sequences for the giants, which have different metallicities between ULySS and FERRE. In the ULySS analysis, there are a number of stars with slightly higher \feh and lower \logg on the hot side of the RGB. In the FERRE analysis, these stars also have lower \logg but similar \feh to the main RGB stars. These stars might be early asymptotic giant branch (AGB) stars. This hypothesis is strengthened by slightly lower carbon abundances for these stars (early AGB stars have lowered carbon abundances due to previous internal mixing processes on the RGB). This observation will be discussed in more detail in future work (Arentsen et al. in prep.).

\begin{figure*}
\centering
\includegraphics[width=0.99\hsize,trim={0.0cm 0.0cm 0.0cm 0.0cm}]{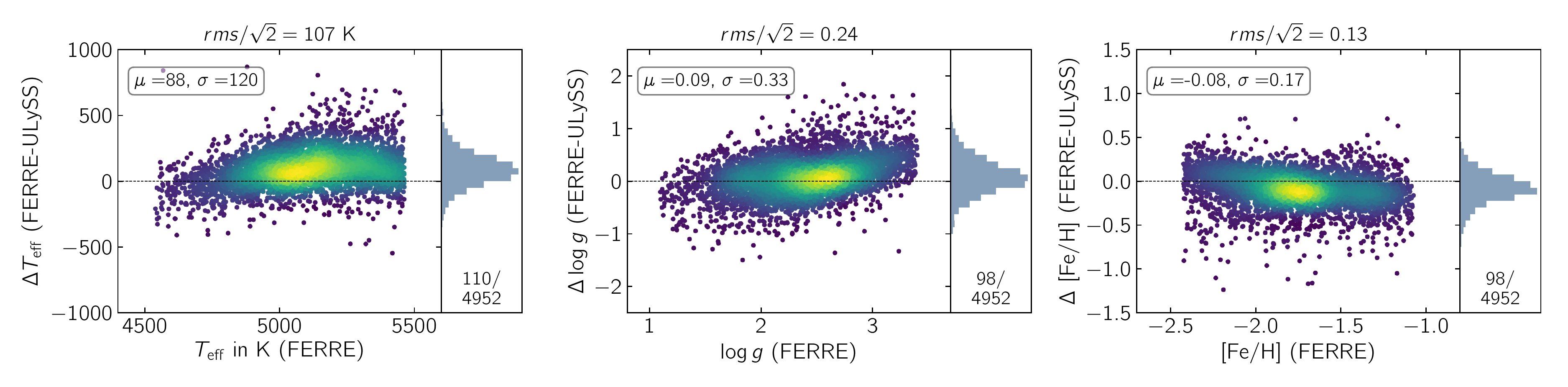}
\caption{Comparison of the parameters between ULySS and FERRE colour-coded by density, for giant stars ($\logg < 3.5$) cooler than 5500~K and with $ -2.5 < \feh < -1.0$. The zero-line is indicated. The mean difference and standard deviation are indicated in the top left corner of each main panel, after discarding 3$\sigma$ outliers. The number of outliers versus the total number of stars are indicated on the bottom of the histogram panels. Finally, the title of each panel includes the corresponding $rms/\sqrt{2}$, which is quadratically added to the internal uncertainties from Section~\ref{sec:ulyssunc} and \ref{sec:ferreunc} to derive total uncertainties for both analyses.}
    \label{fig:comp}
\end{figure*}

\begin{figure*}
\centering
\includegraphics[width=0.9\hsize,trim={0.0cm 0.0cm 0.0cm 0.0cm}]{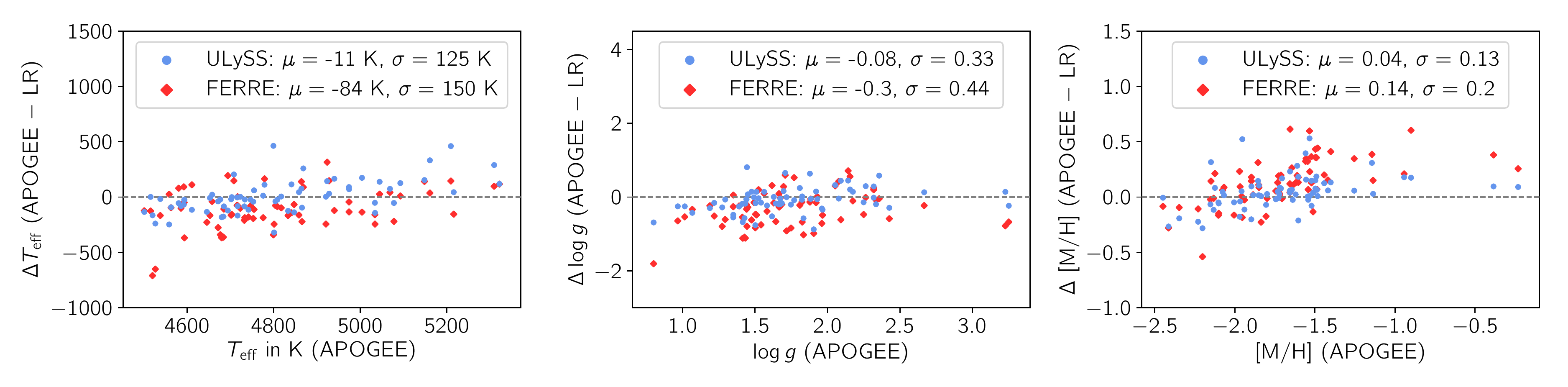}
\caption{Comparison of the parameters between APOGEE and ULySS and FERRE, for cool giants ($\teff < 5500$, $\logg < 3.5$). The mean difference and standard deviation, after discarding 3$\sigma$ outliers, are indicated in the legend of each panel. }
    \label{fig:apogee}
\end{figure*}

\subsection{Total uncertainties}\label{sec:unc}

In Sections~\ref{sec:ulyssunc} and \ref{sec:ferreunc} we provided internal uncertainties for the ULySS and FERRE analyses. These indicate the precision of the derived stellar parameters in both methods. We can use the comparison between ULySS and FERRE to determine the \textit{external} precision, which can be combined with the internal uncertainties to derive total uncertainties. 

The difference in stellar parameters between the two methods is presented in Figure~\ref{fig:comp}, in a region of the parameter space for which both methods are expected to behave well (based on the observations in the previous section). There are some systematic differences between the two analyses. The FERRE temperatures are on average $88$~K higher than those from ULySS, where the discrepancy for the higher temperature stars is slightly larger. The mean difference for \logg is $0.09$~dex higher for FERRE, and it is somewhat larger for the highest \logg stars. Finally, the mean difference in \feh is $0.08$~dex (where FERRE is more metal-poor), varying slightly with \feh. The standard deviations around the mean are $120$~K, $0.33$~dex and $0.17$~dex for \teff, \logg and \feh, respectively. 

A better measure of the external precision than the standard deviation around the mean can be made using the root-mean-square ($rms$), which includes the biases in the deviations from the mean. The external uncertainty can be estimated as $rms/\sqrt{2}$, which assumes that both methods contribute equally to the dispersion. These values have been indicated in the title of each panel in Figure~\ref{fig:comp}, and are $107$~K, $0.24$~dex and $0.13$~dex for \teff, \logg and \feh, respectively (after discarding $3\sigma$ outliers). Total uncertainties can now be derived by quadratically adding the internal ULySS or FERRE uncertainties and the $rms/\sqrt{2}$.  This results in the following median total uncertainties for \teff, \logg and \feh in the ULySS analysis: 111~K, 0.25~dex, and 0.14~dex, respectively, and in the FERRE analysis: 149~K, 0.41~dex and 0.16~dex, respectively. 

The external uncertainties on \cfe cannot be derived in the same way, since it is not a parameter in the ULySS analysis. Instead, they are estimated by scaling the \feh external uncertainties. In the FERRE analysis, the internal uncertainties on \cfe are roughly 1.5 times larger than the \feh internal uncertainties. Assuming that the external uncertainties follow the same trend, an uncertainty floor of $1.5 \times 0.13$~dex~$= 0.20$~dex is added to derive total uncertainties on \cfe. The median of the total uncertainties for \cfe is 0.23~dex.

\subsection{Comparison with APOGEE}
One additional comparison was made with fully external data, by comparing our derived stellar parameters with APOGEE for stars in common between the surveys. Because most bulge surveys contain mainly metal-rich stars, there generally is a lack of stars in common with this work. For instance, in APOGEE DR16 \citep{sdssdr16}, there are only a handful of stars in common with PIGS. This is because APOGEE selects stars ``randomly'' with respect to metallicity, which in the bulge results in a predominantly metal-rich sample. However, we are collaborating with the bulge Cluster APOgee Survey (CAPOS, Geisler et al. in prep.) to obtain APOGEE follow-up of more PIGS stars. To date, one CAPOS field which includes PIGS targets has been fully observed (two additional fields are scheduled to be observed/finished) resulting in a total of 69 stars in common between PIGS and APOGEE DR16+CAPOS with good stellar parameters from both surveys (62 of these come from CAPOS, 7 from DR16). These stars have SNR in APOGEE between 30 and 150, and median uncertainties in \teff, \logg and [M/H] are 134~K, 0.09~dex and 0.03~dex, respectively. 

The AAT observations of the CAPOS field are unfortunately among the worst in the PIGS sample due to relatively bad weather conditions and the sky subtraction is sub-optimal. This affects our FERRE analysis more, because that includes the CaT spectra where the sky subtraction is more important. Additionally, the APOGEE pipeline struggles to provide good stellar parameters for very metal-poor stars \citep[see, e.g., Figure 9 of][which shows that the median absolute error in \feh rises quickly for $\feh < -1.7$, even in very high signal-to-noise spectra, until it reaches about 0.4 dex at $\feh = -2.3$]{Leung19}. The APOGEE-PIGS comparison should therefore not be over-interpreted, but it is still a useful additional indication of our external precision.

The comparison between the PIGS and APOGEE parameters is presented in Figure~\ref{fig:apogee}. The APOGEE [M/H] was used as metallicity, since for many metal-poor stars no \feh is available or uncertain. The dispersion in the parameters between APOGEE and the ULySS analysis is comparable to that of the internal comparison (Figure~\ref{fig:comp}), whereas the dispersions for our FERRE analysis are somewhat larger, as expected due to the lower quality CaT spectra. The biases in the ULySS-APOGEE comparison are negligible, but in the FERRE analysis they are more significant, and in the same direction as in the internal comparison (our FERRE analysis having larger \teff, larger \logg and lower \feh). This could suggest that the biases in the internal comparison are mainly due to FERRE. 

It is important to realise that this comparison with APOGEE is only in a limited range of the parameter space. However, it does show that the typical total uncertainties we have derived from our analysis appear to be realistic. 

\section{Discussion of PIGS survey performance}\label{sec:performance}

\subsection{Metallicity distribution}\label{sec:mdf}

\begin{figure}
\centering
\includegraphics[width=0.8\hsize,trim={0.0cm 0.0cm 0.0cm 0.0cm}]{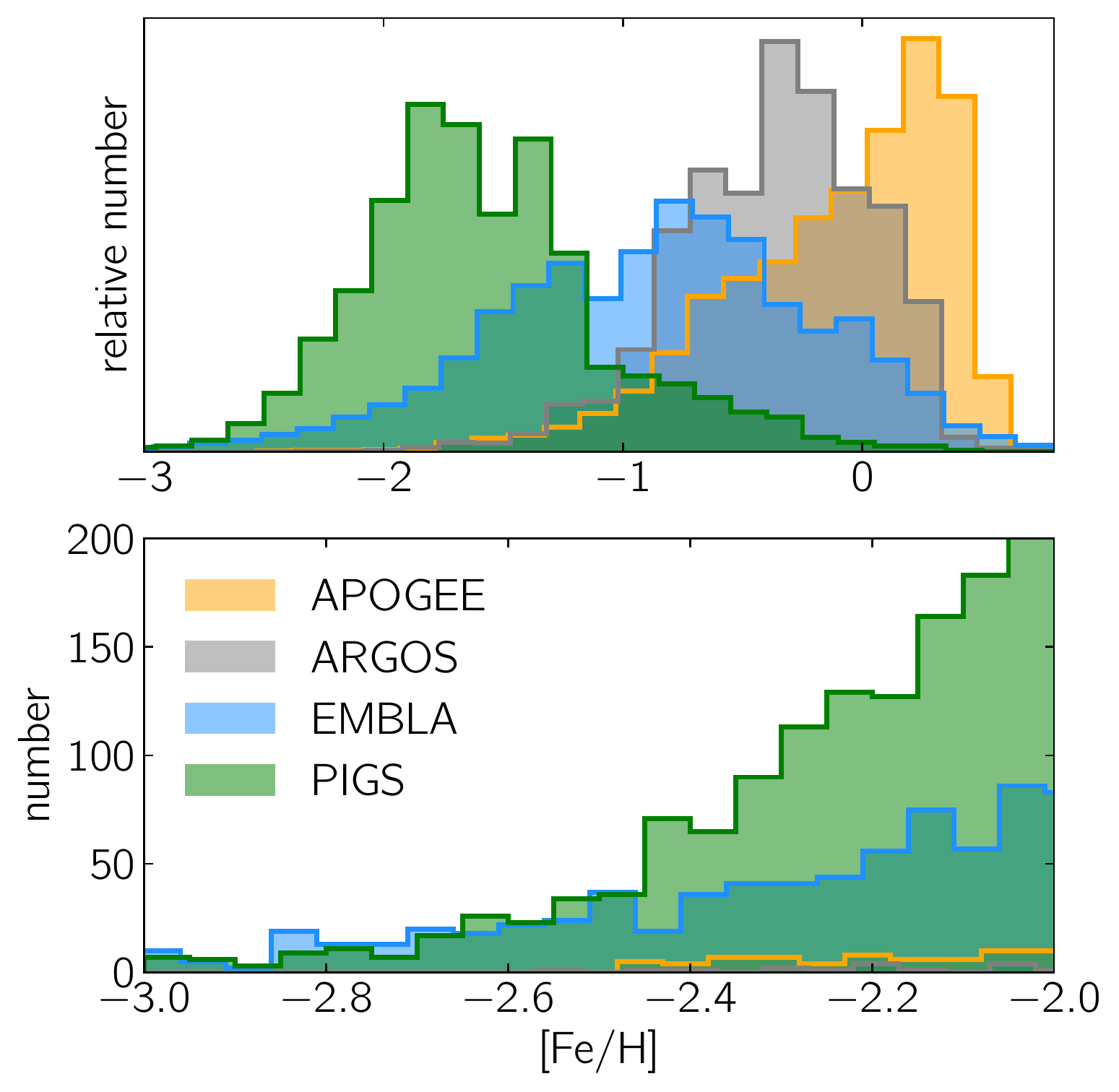}
\caption{Top panel: Relative metallicity distributions of the PIGS, EMBLA, ARGOS and APOGEE inner Galaxy surveys. For PIGS, the FERRE \feh for the AAT sample is shown, with the footprint as in Figure~\ref{fig:footprint}. The EMBLA metallicities are those for their AAT sample (provided by L. Howes, priv. comm.), with a footprint $|l| \lesssim 12^{\circ}$ and $3^{\circ} \lesssim |b| \lesssim 12^{\circ}$. The ARGOS sample consists of all fields with field centres $|l|,|b| \leq 10^{\circ}$ (metallicities provided by M. Ness, priv. comm.). The APOGEE sample shows the [M/H] of all DR16 stars in the region $|l|,|b| < 12^{\circ}$ \citep{sdssdr16}. Bottom panel: absolute number of VMP stars in the same samples.}
    \label{fig:MDF}
\end{figure}

Due to the nature of the PIGS selection, the sample is significantly more metal-poor than typical inner Galaxy surveys like for example ARGOS \citep{ness13a} and APOGEE \citep{sdssdr16}. The PIGS metallicity distribution relative to these two inner Galaxy surveys is presented in the top panel of  Figure~\ref{fig:MDF}. The metallicity distribution from the low-resolution EMBLA survey sample is also included. It is clear that the PIGS sample is unique compared to existing bulge surveys in its large number of low-metallicity stars. 

The PIGS sample contains $\sim1300/160/10$ stars with $\feh < -2.0/-2.5/-3.0$, respectively (from the FERRE analysis). The ten most metal-poor stars are spread between $-3.5 < \feh < -3.0$. This is the largest sample of confirmed very metal-poor (VMP, $\feh < -2.0$) stars towards the inner Galaxy, as shown in the bottom panel of Figure~\ref{fig:MDF}. It is also larger than the sample of VMP stars published by \citet{aguado19} for the main \textit{Pristine} survey, although their sample contains many more stars with $\feh < -2.5$ compared to PIGS. In EMBLA there is a similar number of stars with $\feh < -2.5$ as in PIGS, but the total size of their sample is twice as large as PIGS. There is only one star in common between PIGS and EMBLA with $\feh < -2.5$ and twenty with $\feh < -2.0$, because in most fields we removed the EMBLA stars from our selection. The PIGS stars with $\feh < -2.5$ are distributed relatively homogeneously over the different fields, see the right panel of Figure~\ref{fig:footprint}.

In addition to the VMP stars, the PIGS sample contains $\sim4900$ stars with $-2.0 < \feh < -1.0$, an unprecedentedly large sample of metal-poor inner Galaxy stars to be used to study the metal-poor tail of the Galactic bulge. A first investigation of this population was presented in \citet{arentsen20}, using the metallicities and radial velocities to investigate the rotational signature in the inner Galaxy as a function of metallicity. That work shows that the rotational signal around the Galactic centre decreases with decreasing \feh, while the velocity dispersion increases. Rotation then becomes negligible for the most metal-poor stars, providing a strong constraint for any models attempting to explain the formation and evolution of the inner Galaxy. 

More detailed investigations of this metal-poor sample, including analysis of the orbits of these stars using (future) \Gaia data, are planned for future work. Stars of special interest for dynamical studies are the $\sim1700$ HB stars, for which, being standard candles, accurate distances may be inferred.

\begin{figure*}
\centering
\includegraphics[width=0.8\hsize,trim={0.0cm 0.0cm 0.0cm 0.0cm}]{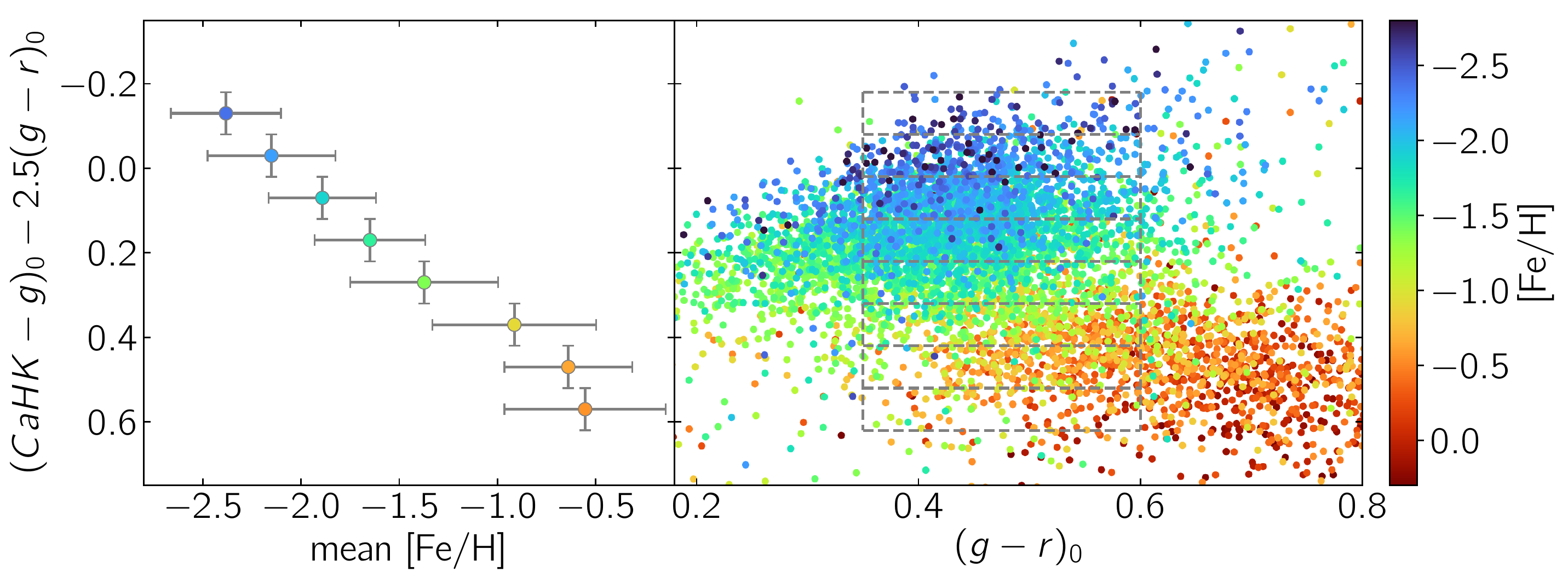}
\caption{Mean metallicity in bins of the y-axis of the Pristine colour-colour diagram (left) and the corresponding diagram with the indicated bins (right). Error bars in the left panel are the 1$\sigma$ dispersion in each bin for \feh, and for the y-axis they indicate the size of the bin. Included are stars from PIGS (using their FERRE \feh), and stars from EMBLA, APOGEE DR16 (using [M/H]) and ARGOS. Only stars with E(B$-$V) $< 0.6$ are included.}
    \label{fig:selectioneff2}
\end{figure*}

\subsection{Survey success rates}\label{sec:photeff}

Using our spectroscopic follow-up sample and spectroscopic observations from the literature, we investigate how well we can produce photometric metallicities in PIGS. For individual stars, photometric metallicities are likely not good, for the reasons described in Section \ref{sec:phot}. Instead, the average metallicity and the metallicity dispersion were determined in sub-regions of the Pristine colour-colour diagram and we investigate how clean a selection can be made for intermediate metal-poor (IMP, $\feh < -1.5$) and VMP (\feh < -2.0) stars. 

In this test, the FERRE spectroscopic metallicities were used instead of those from ULySS, because they are typically better for the VMP stars. We have seen that FERRE is biased for $\feh > -0.5/-1.0$ because the synthetic grid used only includes alpha-enhanced spectra (see Section~\ref{sec:ferre}). However, for the purpose of the tests in this section we included data from three additional spectroscopic surveys that have many more stars with higher metallicities, making the small contribution of metal-rich PIGS stars irrelevant. The three surveys which were added are EMBLA (low-resolution sample, L. Howes priv. comm.), ARGOS (M. Ness priv. comm.) and APOGEE DR16 \citep{sdssdr16}. 

In the right panel of Figure~\ref{fig:selectioneff2} the Pristine colour-colour diagram is shown including all spectroscopic data samples described above. For stars with $0.35 < (g-r)_0 < 0.6$, where the discrimination between different metallicities appears to work best and where most of the stars of interest are located, it is sliced along the y-axis in bins of 0.1~mag. The left-hand panel shows the mean \feh and the dispersion in \feh in each bin. Only stars with E(B$-$V) $ < 0.6$ are used (corresponding to roughly Galactic $b < -4.5^{\circ}$) to avoid strong outliers in very high extinction regions. Additionally, a selection cut using the number of repeated observations and mean flux over error for the $G$-band photometry in \Gaia DR2 was included, to remove stars with variable photometry from the PIGS sample.
 
Figure~\ref{fig:selectioneff2} shows that there is an almost linear relation between the y-axis colour and the metallicity in this regime, which confirms that our box selection for follow-up is a good way of selecting the most metal-poor stars. The \feh dispersion in the bins ranges from $0.30-0.45$~dex, and the typical \feh difference between bins is $\sim 0.2$~dex or more. The largest dispersion is in the intermediate metallicity range, possibly because it contains the largest mixture of stars of different surveys (and each has been analysed differently), or because the metallicity-sensitivity in this regime of the diagram is more non-linear. 

We use the relative number of IMP and VMP stars in our best selection bins to estimate the PIGS selection efficiency for metal-poor stars in the inner Galaxy. 
In the most metal-poor bin, $86\%$ and $80\%$ of the stars have $\feh < -2.0$, for regions with E(B$-$V) $< 0.4$ and $> 0.4$, respectively. If, instead, the two most metal-poor bins are combined the number of stars is increased five-fold (to 500). In these bins, $80\%$ of the stars have $\feh < -2.0$ and $98\%$ have $\feh < -1.5$, for regions with E(B$-$V) $< 0.4$. This goes down to $63\%$ and $93\%$, respectively, in E(B$-$V) $> 0.4$ regions.
These efficiencies do not change significantly when using ($g-i$) as the effective temperature indicator instead of ($g-r$). One might expect the former to work better because the $g$ and $i$ filters cover a longer wavelength baseline, and should thus yield better temperature estimates. However, because they are further apart, the extinction correction of the difference between the two filters is likely less good. These two effects appear to roughly cancel each other in PIGS.

We compare our efficiency to the results from \citet{caseyschlaufman15}, using 2MASS and WISE photometry to identify VMP stars in the bulge region, who report a success rate of 20\% of their candidates having $-3.0 < \feh < -2.0$. The EMBLA survey has not published success rates, but the metallicity distribution of their low-resolution follow-up sample peaks around $\feh = -1.0$ (\citealt{howes16}, see also Figure~\ref{fig:MDF}), whereas PIGS peaks at lower metallicities. Additionally, as discussed in Section \ref{sec:mdf}, their sample has a similar number of stars with \feh $< -2.5$, but twice as many stars in total. We therefore conclude that our selection is more efficient. Taken together, this shows that the efficiency of PIGS for selecting metal-poor stars in the bulge region is unprecedented.

\section{Summary and outlook}\label{sec:conclusion}

In this paper, we have presented the Pristine Inner Galaxy Survey. The $\sim250$ deg$^2$ of photometric $CaHK$ observations in the bulge region, the data reduction and calibration, and the low-/intermediate resolution spectroscopic follow-up of $\sim 8000$ metal-poor candidate stars were described in detail. Robust stellar parameters and metallicities were determined for the spectroscopic data using two completely independent analysis methods: 1) ULySS, a full-spectrum fitting package using an empirical library, and 2) FERRE, a full-spectrum fitting method employing a synthetic stellar model library. Our analysis with ULySS is restricted to the blue arm only, whereas FERRE can fit both blue and red arms of our observations simultaneously. The application of two methods allowed us to make an in-depth comparison between the merits and weaknesses of both methods in the context of analysing metal-poor stars, some of which are inherent to the empirical/synthetic nature of their model spectra. The use of these two complementary methods further enables the estimation of an external uncertainty on top of the internal uncertainties of each of the methods separately. An additional, completely independent, comparison has been carried out with overlapping stars observed with APOGEE. 

As a result of its effective selection of metal-poor targets, PIGS data at the moment includes 1300 stars with spectroscopic $\feh < -2.0$, the largest sample of confirmed very metal-poor stars in the inner Galaxy to date. Combining spectroscopy and photometry, we tested the performance of our photometric selection of metal-poor stars. An efficient selection of $\feh < -2.0$ stars can be made; PIGS has a 90\% success rate using the most conservative selection independent of the extinction (producing only a small sample), and a success rate of 75\%/60\% (intermediate/high extinction) for a sample five times larger. PIGS is therefore the perfect source of metal-poor candidate targets for future spectroscopic surveys in the inner Galaxy, such as for instance the multi-object spectroscopic 4MOST Milky Way Disc and Bulge Low-Resolution Survey \citep[4MIDABLE-LR,][]{chiappini19}. 

Additionally, our sample includes a large sample of metal-poor horizontal branch stars. These stars are interesting in their own right, especially because of their standard candle nature.

Some follow-up work using the sample and analysis presented here is already underway. The third PIGS paper will present carbon abundance results from the FERRE analysis, including an investigation of the sensitivity of the Pristine filter to the carbon abundance, results on the carbon-enhanced metal-poor stars in the inner Galaxy, and an investigation of the evolution of the carbon abundance in metal-poor giant stars. A further paper in the PIGS series will focus in more detail on the dynamical properties of the metal-poor stars in the inner Galaxy using (future) \Gaia data. We plan to derive distances and orbital properties for the spectroscopic metal-poor sample ($\feh < -1.0$). Additionally, as proper motions in the inner Galaxy become more reliable with future Gaia data releases, it may be possible to include the much larger photometric samples of metal-poor stars in kinematic analyses. Finally, we note that there is an ongoing high-resolution spectroscopic follow-up of PIGS stars that will be described elsewhere.

In the coming years, large samples of metal-poor inner Galaxy stars, such as provided through PIGS and other (future) surveys (for example \Gaia DR3+ and 4MOST, \citealt{4most}) will provide a unique view of the central regions of the Milky Way. The chemistry and kinematics of these stars will lead to insights into the early history of the Milky Way, and large samples of them will allow for the discovery of rare stars whose abundance patterns teach us about previous, early generations of stars, all the way back to the First Stars. 

\section*{Acknowledgements}

We thank Friedrich Anders for discussions on the extinction in \Gaia, and for sharing his \Gaia extinction coefficients. We thank Ting S. Li for valuable discussions and efforts regarding the fibre throughput issues of 2dF. We thank Louise Howes and Melissa Ness for sharing their spectroscopic catalogues with us, this was very helpful in the early development of PIGS and for this paper.

We thank the Australian Astronomical Observatory, which have made these observations possible. We acknowledge the traditional owners of the land on which the AAT stands, the Gamilaraay people, and pay our respects to elders past and present. Based on data obtained at Siding Spring Observatory (via programs S/2017B/01, A/2018A/01, OPTICON 2018B/029 and OPTICON 2019A/045, PI: A. Arentsen). 
	
Based on observations obtained with MegaPrime/MegaCam, a joint project of CFHT and CEA/DAPNIA, at the Canada-France-Hawaii Telescope (CFHT) which is operated by the National Research Council (NRC) of Canada, the Institut National des Science de l'Univers of the Centre National de la Recherche Scientifique (CNRS) of France, and the University of Hawaii.
	
Based on observations obtained through the Chilean National Telescope Allocation Committee through programs CN2017B-37, CN2018A-20, CN2018B-46, CN2019A-98 and CN2019B-31.
	
AA, ES and KY gratefully acknowledge funding by the Emmy Noether program from the Deutsche Forschungsgemeinschaft (DFG).
NFM gratefully acknowledges support from the French National Research Agency (ANR) funded project ``Pristine'' (ANR-18-CE31-0017) along with funding from CNRS/INSU through the Programme National Galaxies et Cosmologie and through the CNRS grant PICS07708.
DA thanks the Leverhulme Trust for financial support.
KAV  thank  the  Natural  Sciences  and  Engineering Research Council for funding through the Discovery Grants program and the CREATE training program on New Technologies for Canadian Observatories.
DBZ acknowledges the support of the Australian Research Council through Discovery Project grant DP180101791.
CAP and JIGH acknowledge financial support from the Spanish Ministry of Science and Innovation (MICIIN) under the project MICIIN AYA2017-86389-P. JIGH also acknowledges the Spanish MICIIN under 2013 Ram\'on y Cajal program MICIIN RYC-2013-14875.
D.G. gratefully acknowledges support from the Chilean Centro de Excelencia en Astrof\'isicay Tecnolog\'ias Afines (CATA) BASAL grant AFB-170002. D.G. also acknowledges financial support from the Direcci\'on de Investigaci\'on y Desarrollo de la Universidad de La Serena through the Programa de Incentivo a la Investigaci\'on de Acad\'emicos (PIA-DIDULS).
Horizon 2020: This project has received funding from the European Union's Horizon 2020 research and innovation programme under grant agreement No 730890. This material reflects only the authors views and the Commission is not liable for any use that may be made of the information contained therein.

The authors thank the International Space Science Institute, Bern, Switzerland for providing financial support and meeting facilities to the international team "Pristine". 
	
This work has made use of data from the European Space Agency (ESA) mission {\it Gaia} (\url{https://www.cosmos.esa.int/gaia}), processed by the {\it Gaia} Data Processing and Analysis Consortium (DPAC, \url{https://www.cosmos.esa.int/web/gaia/dpac/consortium}). Funding for the DPAC has been provided by national institutions, in particular the institutions participating in the {\it Gaia} Multilateral Agreement. 

The Pan-STARRS1 Surveys (PS1) and the PS1 public science archive have been made possible through contributions by the Institute for Astronomy, the University of Hawaii, the Pan-STARRS Project Office, the Max-Planck Society and its participating institutes, the Max Planck Institute for Astronomy, Heidelberg and the Max Planck Institute for Extraterrestrial Physics, Garching, The Johns Hopkins University, Durham University, the University of Edinburgh, the Queen's University Belfast, the Harvard-Smithsonian Center for Astrophysics, the Las Cumbres Observatory Global Telescope Network Incorporated, the National Central University of Taiwan, the Space Telescope Science Institute, the National Aeronautics and Space Administration under Grant No. NNX08AR22G issued through the Planetary Science Division of the NASA Science Mission Directorate, the National Science Foundation Grant No. AST-1238877, the University of Maryland, Eotvos Lorand University (ELTE), the Los Alamos National Laboratory, and the Gordon and Betty Moore Foundation.
	
This publication makes use of data products from the Two Micron All Sky Survey, which is a joint project of the University of Massachusetts and the Infrared Processing and Analysis Center/California Institute of Technology, funded by the National Aeronautics and Space Administration and the National Science Foundation.

This research has made use of "Aladin sky atlas" developed at CDS, Strasbourg Observatory, France \citep{aladin1}. It also made extensive use of the \textsc{matplotlib} \citep{matplotlib}, \textsc{pandas} \citep{pandas}, \textsc{astropy} \citep{astropy13,astropy18} and \textsc{dustmaps} \citep{dustmaps} Python packages, and of \textsc{Topcat} \citep{topcat}.


\bibliographystyle{mnras}
\bibliography{mpbulgeII.bib}   


\section*{Affiliations}

\begin{small}
\textit{
\noindent $^{1}$Leibniz-Institut f\"ur Astrophysik Potsdam (AIP), An der Sternwarte 16, D-14482 Potsdam, Germany\\
$^{2}$Universit\'e de Strasbourg, CNRS, Observatoire astronomique de Strasbourg, UMR 7550, F-67000 Strasbourg, France\\
$^{3}$Max-Planck-Institut f\"ur Astronomie, K\"onigstuhl 17, D-69117 Heidelberg, Germany\\
$^{4}$Institute of Astronomy, University of Cambridge, Madingley Road, Cambridge CB3 0HA, UK \\
$^{5}$Department of Physics and Astronomy, Macquarie University, Sydney, NSW 2109, Australia \\
$^{6}$Instituto de Astrof\'isica de Canarias, V\'ia L\'actea, 38205 La Laguna, Tenerife, Spain \\
$^{7}$Universidad de La Laguna, Departamento de Astrof\'isica, 38206 La Laguna, Tenerife, Spain \\
$^{8}$Universit\'e C\^ote d'Azur, Observatoire de la C\^ote d'Azur, CNRS, Laboratoire Lagrange, Blvd de l'Observatoire, F-06304 Nice, France \\
$^{9}$Department of Physics \& Astronomy, University of Victoria, Victoria, BC, V8W 3P2, Canada \\
$^{10}$Department of Astronomy \& Astrophysics, University of Toronto, Toronto, ON M5S 3H4, Canada \\
$^{11}$Institute of Astronomy of the Russian Academy of Sciences, Pyatnitskaya st. 48, 119017, Moscow, Russia \\
$^{12}$UK Astronomy Technology Centre, Royal Observatory Edinburgh, Blackford Hill, Edinburgh, EH9 3HJ, UK \\
$^{13}$NRC Herzberg Astronomy and Astrophysics, 5071 West Saanich Road, Victoria, BC, V9E 2E7, Canada \\
$^{14}$Sydney Institute for Astronomy, School of Physics, A28, The University of Sydney, NSW 2006, Australia \\
$^{15}$School of Physics, UNSW, Sydney, NSW 2052, Australia \\
$^{16}$Space Telescope Science Institute, 3700 San Martin Drive, Baltimore, MD 21218, USA \\
$^{17}$Departamento de Astronom{\'{\i}}a, Universidad de Concepci{\'o}n, Casilla 160-C, Concepci{\'o}n, Chile \\
$^{18}$Instituto de Investigaci\'on Multidisciplinario en Ciencia y Tecnolog\'ia, Universidad de La Serena, Avenida Ra\'ul Bitr\'an S/N, La Serena, Chile \\
$^{19}$Departamento de Astronom\'ia, Facultad de Ciencias, Universidad de La Serena, Av. Juan Cisternas 1200, La Serena, Chile
}
\end{small}

\bsp	
\label{lastpage}
\end{document}